\documentclass[12pt]{article}

\usepackage{setspace}
\usepackage{amsmath,bm}
\usepackage{amsmath,empheq}
\usepackage{amsfonts}
\usepackage{graphicx}
\usepackage{cases}
\usepackage[font=small]{caption}
\usepackage{subcaption}
\usepackage{amsmath}
\usepackage{enumerate}
\usepackage[table]{xcolor}
\usepackage{verbatim}

\graphicspath{{Figures/}}
\usepackage[numbers]{natbib}
\usepackage{csquotes}
\usepackage{makecell}
\usepackage{rotating}
\usepackage{multirow}
\usepackage{enumitem}
\usepackage{tabularx, booktabs}
\usepackage{textgreek}
\newcolumntype{Y}{>{\centering\arraybackslash}X}
\usepackage{footnote}
\usepackage[bottom]{footmisc}
\makesavenoteenv{tabular}
\makesavenoteenv{table}
\usepackage[export]{adjustbox}
\usepackage{mathrsfs}
\usepackage{color,soul}
\usepackage{booktabs}
\usepackage{floatrow}
\usepackage{tikz}
\usetikzlibrary{decorations.pathreplacing,decorations.pathmorphing}
\usepackage{setspace}
\newfloatcommand{capbtabbox}{table}[][\FBwidth]
\usepackage[toc,page]{appendix}
\usepackage{tikz}
\DeclareRobustCommand\sampleline[1]{%
  \tikz\draw[#1] (0,0) (0,\the\dimexpr\fontdimen22\textfont2\relax)
  -- (2em,\the\dimexpr\fontdimen22\textfont2\relax);%
}
\definecolor{ao(english)}{rgb}{0.0, 0.5, 0.0}
\definecolor{alizarin}{rgb}{0.82, 0.1, 0.26}
\definecolor{aqua}{rgb}{0.0, 0.6, 1.0}
\definecolor{ao}{rgb}{0.0, 0.0, 1.0}

\newcommand{\bs}[1]{\boldsymbol{#1}}
\usepackage{siunitx}
\AtBeginDocument{\RenewCommandCopy\qty\SI}
\usepackage{textcomp}
\usepackage{hyperref}
\usepackage[nameinlink]{cleveref}
\usepackage{soul}

\usepackage{bigints}
\usepackage{mwe}
\usepackage{algorithm,algpseudocode}
\usepackage{relsize}

\newcounter{algsubstate}

\newenvironment{algsubstates}
  {\setcounter{algsubstate}{0}%
   \renewcommand{\State}{%
     \refstepcounter{algsubstate}%
     \Statex {\footnotesize\alph{algsubstate}:}\space}}
  {}

\algnewcommand{\algorithmicgoto}{\textbf{go to}}%
\algnewcommand{\Goto}[1]{\algorithmicgoto~\ref{#1}}%

\usepackage{amsmath}
\usepackage{amsbsy}
\usepackage{amssymb}
\usepackage{amscd}
\usepackage{amsfonts}

\newcommand{\bfd}{{\mathbf{d}}}
\newcommand{\bfe}{{\mathbf{e}}}
\newcommand{\bff}{{\mathbf{f}}}

\newcommand{\bfu}{{\mathbf{u}}}
\newcommand{\bfv}{{\mathbf{v}}}

\newcommand{\bfB}{{\mathbf{B}}}

\newcommand{\bfD}{{\mathbf{D}}}

\newcommand{\bfF}{{\mathbf{F}}}

\newcommand{\bfK}{{\mathbf{K}}}
\newcommand{\bfL}{{\mathbf{L}}}
\newcommand{\bfM}{{\mathbf{M}}}
\newcommand{\bfN}{{\mathbf{N}}}

\newcommand{\bfP}{{\mathbf{P}}}

\newcommand{\bfT}{{\mathbf{T}}}

\newcommand{\bfX}{{\mathbf{X}}}

\newcommand{\beq}{\begin{equation}}
\newcommand{\eeq}{\end{equation}}
\newcommand{\beqs}{\begin{eqnarray}}
\newcommand{\eeqs}{\end{eqnarray}}
\newcommand{\beql}{\begin{equation} \label}

\newcommand{\calH}{{\cal H}}

\newcommand{\calM}{{\cal M}}

\newcommand{\calV}{{\cal V}}
\newcommand{\calW}{{\cal W}}

\newcommand{\ttrace}{\mathop{\rm tr}\nolimits}

\begin{document}

\title{\Large Variational multiscale enrichment method for dynamic response of hyperelastic materials at finite deformation}

\author{
\large{ Abhishek Arora$^{\ddagger}$
and Caglar Oskay$^{\ddagger\S}$\footnote{Corresponding author address: VU Station B\#351831, 2301 Vanderbilt Place, Nashville, TN 37235. Email: caglar.oskay@vanderbilt.edu}} \\ 
	\\
	{$^\ddagger$\large Department of Civil and Environmental Engineering} \\ 
	{$^\S$\large Department of Mechanical Engineering} \\ 
	{\large Vanderbilt University} \\
	{\large Nashville, TN 37212} 
}
\date{}
\maketitle
\setstretch{1.3}


\abstract{In this manuscript, we extend the variational multiscale enrichment (VME) method to model the dynamic response of hyperelastic materials undergoing large deformations. This approach enables the simulation of wave propagation under scale-inseparable conditions, including short-wavelength regimes, while accounting for material and geometric nonlinearities that lead to wave steepening or flattening. By employing an additive decomposition of the displacement field, we derive multiscale governing equations for the coarse- and fine-scale problems, which naturally incorporate micro-inertial effects. The framework allows the discretization of each unit cell with a patch of coarse-scale elements, which is essential to accurately capture wave propagation in short-wavelength regimes. An operator-split procedure is used to iteratively solve the semi-discrete equations at both scales until convergence is achieved. The coarse-scale problem is integrated explicitly, while the fine-scale problem is solved using either explicit or implicit time integration schemes, including both dissipative and non-dissipative methods. Numerical examples demonstrate that multiscale dissipative schemes effectively suppress spurious oscillations. The multiscale framework was applied to investigate how material and geometric nonlinearities, along with elastic stiffness contrast in heterogeneous microstructures, influence key wave characteristics such as dispersion, attenuation, and steepening. This multiscale computational framework provides a foundation for studying the dynamic response of architected materials.

\vspace{.3cm}
}
%
%
\vspace{-0.3cm}
\section{Introduction} \label{sec:introduction}
Architected materials have attracted widespread interest in the research community due to their ability to achieve exceptional properties by tailoring geometric features across multiple length scales, often outperforming conventional materials in terms of mechanical, thermal, or functional performance. Energy-absorbing architected materials such as honeycombs and auxetic metamaterials demonstrate favorable impact resistance and deformation behavior, making them excellent candidates for crash protection and vibration damping applications (see e.g., \cite{gibson2010cellular,zheng2014ultralight, yuan20193d, francisco2022review, bian2024design}). Computational simulation of the transient dynamic response of structures composed of architected unit cells using direct numerical simulations (DNS) is computationally expensive and cumbersome, especially in scenarios involving large structural domains or complex microstructural features. As a result, there is a critical need to develop multiscale modeling techniques that can efficiently simulate the dynamic response of such structures while preserving accuracy.

Various homogenization approaches have been proposed to model the dynamic response of heterogeneous materials based on the assumption that the microstructural size is much smaller than the wavelength, commonly referred to as the scale separation limit. In the computational homogenization approach \cite{pham2013transient, bensoussan2011asymptotic, liu2017variational, roca2018computational}, the fundamental idea is to characterize the material response locally at each quadrature point of the finite element discretization of the macroscopic domain by the numerical evaluation of a representative volume element (RVE) or a periodic unit cell. Nested initial-boundary value problems at the macro- and microscales are evaluated, with scale bridging relations that satisfy energy consistency between scales, i.e., the Hill-Mandel condition. Higher-order asymptotic homogenization approaches were developed to capture the dynamic homogenized response at the short-wavelength regime \cite{andrianov2013dynamic, fish2002non, hui2014high, hu2017nonlocal, hu2018spatial, hu2019multiscale}. Some alternative multiscale methods that do not rely on the assumption of scale separation, including the elastodynamic homogenization methods based on Willis' theory \cite{willis1997dynamics, milton2007modifications, nemat2011overall, nassar2016generalized, meng2018dynamic}, multiscale finite element method \cite{casadei2013geometric, casadei2016multiscale}, and the method of computational continua \cite{fish2010computational, fafalis2015computational} have been used for wave propagation in short-wavelength regime. The computational homogenization framework has also been extended to lattice metamaterials, including both truss- and beam-based architectures \cite{glaesener2019continuum, glaesener2020continuum, le2023numerical}, though most developments to date have primarily focused on quasistatic loading conditions.

The variational multiscale method (VMM)~\cite{hughes1998variational} is an alternative multiscale strategy that does not assume scale separation. This method is based on the additive split of the cardinal response field into coarse and fine-scale components, resulting in separate but coupled multiscale problems. Numerical efficiency is achieved by evaluating the fine-scale problem analytically (when an analytical form that approximates fine-scale response is known; e.g., see \cite{oberai1998multiscale}). Problems that involve complex micro-morphologies and nonlinearities may not have analytical forms to represent the fine-scale response with sufficient accuracy. This prompted VMM variants that treat the fine-scale problem numerically, such as the numerical subgrid upscaling method~\cite{arbogast2002implementation}, the stochastic variational multiscale method~\cite{asokan2006stochastic, ganapathysubramanian2007modeling}, and the variational multiscale enrichment method~\cite{oskay2012variational, oskay2013variational, zhang2015variational, zhang2016reduced, zhang2017reduced}. Of particular relevance is the spectral variational multiscale enrichment \cite{hu2020spectral}, which was developed to model the transient dynamics and wave propagation of phononic crystals and acoustic metamaterials. This study focused on the linear material behavior, and the effects of geometric and material nonlinearities on wave propagation were not considered. 

Obtaining accurate solutions for transient dynamics or wave propagation problems in the short-wavelength regime (using either direct numerical simulations or multiscale methods) is a challenge. Numerical errors due to spatial and time discretizations using finite element methods and time integration schemes lead to artificial period elongations and amplitude decays, which manifest themselves as numerical dispersion and dissipation errors \cite{bathe2006finite, noh2013performance}. A plethora of approaches have been developed to address the dispersion and dissipation errors, including the use of higher-order spatial discretizations~\cite{gottlieb1977numerical, gamallo2006partition}, finite element interpolations enriched with wave packets for multiscale wave propagation problems~\cite{kohno2010finite}, or the use of spectral elements in the context of multiscale description~\cite{hu2020spectral}.

Higher-order elements can improve accuracy but are often too expensive, motivating the use of time integration schemes with lower-order elements that incorporate controlled numerical dissipation~\cite{bathe2006finite, kuhl1999energy, fung2003numerical}. The Bathe implicit method~\cite{bathe2005composite} addresses this by filtering out unresolved high-frequency modes while accurately integrating the resolvable ones, thereby reducing dispersion errors. Extending this idea, Noh and Bathe~\cite{noh2013explicit} proposed an explicit scheme with high-frequency dissipation that preserves second-order accuracy and produces the desired behavior for period elongations and amplitude decays, small for small time steps and rapidly increasing for larger ones. In contrast, the Tchamwa--Wielgosz scheme~\cite{tchamwa1999accurate}, though only first-order accurate, performs better than many classical explicit schemes~\cite{newmark1959method, chung1994new, zhai1996two, nsiampa2008comparative}, but is still outperformed by Noh and Bathe’s method.

In this study, we present a robust multiscale simulation framework based on the variational multiscale enrichment (VME) method for modeling the transient dynamic response of hyperelastic composite materials undergoing large deformations. The approach employs an additive decomposition of displacement, velocity, and acceleration fields into coarse- and fine-scale components, enabling a consistent derivation of the multiscale governing equations within a Lagrangian setting. To accurately resolve wave propagation in short-wavelength regimes, the formulation allows each unit cell to be discretized into multiple coarse-scale elements. The coupled coarse- and fine-scale problems are solved iteratively using an operator-split procedure until convergence is achieved. 

The semi-discrete multiscale equations are integrated explicitly using either the non-dissipative central difference method or the dissipative explicit scheme of Noh and Bathe~\cite{noh2013explicit}. Although the VME formulation is designed for scale-inseparable problems, differences in characteristic time scales between coarse and fine discretizations can arise. In such cases, particularly when accuracy requirements permit larger time steps than the stability limits of the fine-scale problem, a mixed integration strategy can be adopted -- explicit integration for the coarse-scale problem and implicit integration of the fine-scale problem, for example, with the Bathe implicit method~\cite{bathe2005composite}. Numerical investigations show that spurious oscillations appear in both direct numerical and multiscale simulations when using the central difference method, but these are effectively suppressed by dissipative integration schemes. The proposed multiscale framework is shown to capture wave dispersion, attenuation, and wave steepening driven by microstructural heterogeneity, unit cell size, and geometric and material nonlinearities. Finally, the computational performance of different time integration schemes is assessed for microstructures with varying elastic modulus contrasts.

The remainder of this manuscript is organized as follows: Section~\ref{sec:VME_Formulation} presents the formulation of the multiscale method. Section~\ref{sec:computational_formulation_two_scale_pdes} provides the computational approximation of the resulting coarse- and fine-scale partial differential equations (PDEs) from the VME method, along with the time integration schemes for the multiscale problem. Further details about the evaluation of non-standard element-level matrices and vectors in multiscale discretization, and estimation of stable time increment for the multiscale problem are discussed. Section~\ref{sec:numerical_verification} provides examples of wave propagation in 1-D problems with homogeneous and heterogeneous microstructures.  Section~\ref{sec:conclusion} discusses concluding remarks and future research directions.
%
%
\vspace{-0.3cm}
\section{Variational Multiscale Enrichment Formulation} \label{sec:VME_Formulation}
In this section, we introduce the formulation of the proposed VME approach for modeling the dynamic response of hyperelastic materials. Let us denote a heterogeneous body with $\Omega \in \mathbb{R}^{n_{\mathrm{sd}}}$ ($n_{\mathrm{sd}}$=1,2 or 3), an open and bounded domain composed of repeated unit cells. The governing equations for linear and angular momentum balance in the Lagrangian setting are as follows:
\begin{subequations}
\begin{align}
    & \bs{\nabla}_{\bfX} \cdot \bfP (\bfX , t)  + \bfB (\bfX , t) = \rho_0 \: \ddot{\bfu} (\bfX , t); \quad \bfX \in \Omega, \quad t \in [0,T] \label{eq:lmb}\\
    & \bfP (\bfX , t) \: \bfF^T (\bfX , t) = \bfF(\bfX , t) \:  \bfP^T(\bfX , t); \quad \bfX \in \Omega, \quad t \in [0,T] \label{eq:amb}
\end{align}
\end{subequations}
where $\bfX$ denotes the Cartesian coordinates in the reference (undeformed) configuration, $t$ the time coordinate, $\bfP$ the First Piola-Kirchhoff stress, $\bfF$ the deformation gradient, $\bfB$ the body force, $\rho_0$ the mass density in the reference configuration, and $\bfu$ the displacement field. $\bs{\nabla}_{\bfX} \cdot$ denotes the divergence operator in the reference configuration, and $\ddot{(\cdot)}$ denotes the second time derivative. 

The Dirichlet and Neumann boundary conditions are specified respectively on $\Gamma^u \subseteq \partial \Omega $ and $\Gamma^t \subseteq \partial \Omega$, such that $\Gamma^u \cup \Gamma^t = \partial \Omega$ and $\Gamma^u \cap \Gamma^t = \varnothing$:
\begin{subequations}
\begin{align}
     & \bfu (\bfX, t) = \tilde{\bfu} (\bfX, t); \quad \bfX \in \Gamma^u \\
    & \bfP (\bfX, t) \cdot \bfN = \Tilde{\bfT} (\bfX, t);  \quad \bfX \in \Gamma^t  
\end{align}    
\end{subequations}
where $\bfN$ denotes the outward unit normal in the reference configuration, $\tilde{\bfu}$ and $\tilde{\bfT}$ are the prescribed displacement and traction vectors, respectively. The initial conditions are specified as follows:
\begin{subequations} \label{eq:bcs}
\begin{align}
    & \bfu (\bfX, 0) = \hat{\bfu} (\bfX); \quad \bfX \in \Omega \\
    & \dot{\bfu} (\bfX, 0) = \hat{\bfv} (\bfX); \quad  \bfX \in \Omega
\end{align}    
\end{subequations}
in which $\hat{\bfu}$ and $\hat{\bfv}$ denote the prescribed displacement and velocity fields respectively at $t=0$.

At any given material point, $\bfX$, the constituent material is taken to follow a hyperelastic constitutive law. The first Piola-Kirchhoff stress tensor is given by:
\begin{equation}
    \bfP = \frac{\partial \psi (\bfF)}{ \partial \bfF}, 
\end{equation}
where $\psi$ denotes the strain energy density function. The angular momentum balance given by Eq.~\eqref{eq:amb} is satisfied by the objectivity of the strain energy density function.

The weak form of the linear momentum balance (Eq.~\eqref{eq:lmb})  with the boundary conditions in Eq.~\eqref{eq:bcs} is:
\begin{equation} \label{eq:weak_form_onescale}
    \int_{\Omega}  \bs{\nabla}_{\bfX} \delta \bfu : \bfP \: dV + \int_{\Omega} \rho_0  \delta \bfu \cdot \ddot{\bfu} \: dV = \int_{\Omega} \delta \bfu \cdot \bfB \: dV + \int_{\Gamma^t} \delta \bfu \cdot \Tilde{\bfT}  \: dA, 
\end{equation}
where $\delta \bfu$ denotes the test function. The function spaces for the trial and test functions are, respectively:
\begin{subequations}
\begin{align}
    & \calV = \left\{ \bfu \: \vert \: \bfu \in H^1(\Omega), \: \bfu = \tilde{\bfu} \: \: \textrm{on} \: \Gamma^u  \right\}, \\
    & \calW = \left\{ \delta \bfu \: \vert \: \delta \bfu \in H^1(\Omega), \: \delta \bfu = \mathbf{0} \: \: \textrm{on} \: \Gamma^u  \right\},
\end{align}    
\end{subequations}
where $H^1$ is the Sobolev space consisting of functions whose values and first weak derivatives are square-integrable.

We make some assumptions about the problem domain and discretization: The geometry of the domain $\Omega$ can be partitioned into $n_{\mathrm{es}}$ non-overlapping identically shaped enrichment subdomains. Each enrichment subdomain is associated with a unit cell and discretized using a coarse patch of elements, and the number of coarse-scale finite elements in a patch is denoted by $n_{\mathrm{ecp}}$. The total number of coarse-scale elements is given by $n_{\mathrm{ec}}= n_{\mathrm{es}} \times n_\mathrm{ecp}$.  The interior of a subdomain, $\alpha$, is denoted as ${\Omega}_{\alpha}$ and its boundary is denoted as $\Gamma_{\alpha}$, with the overall domain partitioning into subdomains is performed such that  $\overline{\Omega} = \cup_{\alpha=1}^{n_\mathrm{es}} \: \overline{\Omega}_{\alpha}$ with $\overline{(\cdot)}$ represents the closure of $(\cdot)$. In addition to the coarse-scale discretization, each subdomain is separately discretized using $n_{\mathrm{ef}}$ fine-scale elements that resolve the features of the underlying unit cell. 

The displacement field over the problem domain is decomposed into coarse and fine-scale contributions using a two-scale additive decomposition:
\begin{equation} \label{eq:disp_split}
    \bfu  = \bfu^\mathrm{c} + \sum_{\alpha = 1}^{ n_\mathrm{es}} \calH (\Omega_{\alpha}) \bfu^{\mathrm{f} , \alpha} ,
\end{equation}
where $\calH(\cdot)$ is an indicator function defined below:
\begin{equation} \label{eq:inidicator_function}
\calH (\Omega_{\alpha}) = \left\{ 
\begin{aligned}
    & 1, \quad  \mathrm{for} \: \bfX \in \Omega_{\alpha} \\ 
    & 0, \quad \mathrm{elsewhere}.                      
\end{aligned}
\right.
\end{equation}
The indicator function in Eq.~\eqref{eq:inidicator_function} ensures that only the fine-scale response associated with subdomain $\Omega_{\alpha}$ contributes to the displacement field within the subdomain. The coarse-scale field captures the slowly varying component of the solution, whereas the fine-scale fields resolve the rapidly varying solution due to material heterogeneity. We note that Eq.~\ref{eq:disp_split} does not directly satisfy the continuity condition on $\bfu$. The continuity is satisfied by appropriately selecting the boundary conditions for the fine-scale response field~\cite{oskay2012variational}. The test function is decomposed similarly:
\begin{equation} \label{eq:variational_disp_split}
    \delta \bfu = \delta \bfu^\mathrm{c} + \sum_{\alpha = 1}^{ n_\mathrm{es}} \calH (\Omega_{\alpha}) \: \delta \bfu^{\mathrm{f}, \alpha}.
\end{equation}

The finite-dimensional subspaces (following the finite dimensional approximation consistent with the standard finite element method) for the coarse-scale trial and test functions are denoted as $\calV^\mathrm{c}$ and $\calW^\mathrm{c}$, respectively, and the corresponding fine-scale function spaces are $\calV_{\alpha}^\mathrm{f}$ and $\calW_{\alpha}^\mathrm{f}$. These spaces are selected such that their direct sum forms the finite-dimensional subspaces for the original (single-scale) trial and test functions:
\begin{equation} \label{directSum}
\begin{aligned}
     \calV^h = \calV^\mathrm{c}  \oplus \bigoplus_{\alpha = 1}^{n_{\mathrm{es}}} \calV_{\alpha}^{\mathrm{f}}, \quad  \calW^h = \calW^\mathrm{c} \oplus \bigoplus_{\alpha = 1}^{n_{\mathrm{es}}} \calW_{\alpha}^{\mathrm{f}}. 
\end{aligned}
\end{equation}

The fine-scale spaces are defined ($\calW_{\alpha}^\mathrm{f} = \calV_{\alpha}^\mathrm{f}$) such that the trial and test functions are non-zero only within the corresponding enrichment subdomain $\Omega_{\alpha}$ and vanish elsewhere except at the subdomain boundary where the external traction (denoted by $\Gamma^t_{\alpha}$) is applied:
\begin{equation} \label{fineScaleFunDef}
\calV_{\alpha}^\mathrm{f} \ni \bfu = \mathbf{0} ; \: \: \textrm{when} \: \: \bfX \in \Gamma^u_{\alpha},
\end{equation}
%
where $\Gamma^u_{\alpha} = \Gamma_{\alpha} \backslash \Gamma^t_{\alpha}$. Equation \eqref{fineScaleFunDef} implies that homogeneous Dirichlet boundary conditions are imposed at the boundary of each subdomain $\Omega_{\alpha}$ for fine-scale displacement fields, except at $\Gamma^t_{\alpha}$ boundary where the external traction is applied. This boundary condition has been previously used in similar work \citep{hu2020spectral} for modeling transient dynamical response of phononic crystals and acoustic metamaterials and in other works \citep{oskay2012variational, zhang2016reduced, masud2004multiscale, masud2006variational}. Other boundary conditions (e.g., mixed boundary conditions) have also been previously investigated~\cite{zhang2015variational}. Moreover, it follows from the direct sum decomposition given in Eq.~\eqref{directSum} that $\calV^\mathrm{f}_{\alpha} \cap \calV^\mathrm{f}_{\beta} = \varnothing$. Similarly, in order to ensure direct sum decomposition, the finite-dimensional coarse-scale function spaces $\calV^\mathrm{c} \subset \calV$ and $\calW^\mathrm{c} \subset \calW$ are selected such that:
\begin{equation}
\left\| \bfu - \mathbf{v} \right\|_{\Omega_\alpha} \ne \mathbf{0}; \hspace{0.3cm} \bfu \in \calV^\mathrm{c}; \,\,\, \mathbf{v} \in \calV_{\alpha}^\mathrm{f},
\end{equation}
for any ($\bfu$,$\mathbf{v}$) pair, and $\| \cdot \|_{\Omega_\alpha}$ is L$^{2}$ norm over ${\Omega_\alpha}$.



Substituting Eqs.~\eqref{eq:disp_split} and \eqref{eq:variational_disp_split} in Eq.~\eqref{eq:weak_form_onescale}, one can decompose the weak form of the linear momentum balance equation into two tightly coupled problems. The  coarse-scale problem is defined over the entire problem domain as follows:
\begin{equation}\label{eq:coarse-scale-weakform}
\begin{multlined}
      \int_{\Omega} \delta \bfu^\mathrm{c} \cdot \rho_0 \:  \ddot{\bfu}^\mathrm{c}  \: dV  + \int_{\Omega} \boldsymbol{\nabla}_{\bfX} \delta \bfu^\mathrm{c} : \bfP \left(\bfX, t ,  \bfu^\mathrm{c}, {\color{red}\bfu^{\mathrm{f} ,\alpha}} \right) \: dV = \hspace{4.7cm} \\ 
    \quad - \sum_{\alpha = 1}^{ n_\mathrm{es}} \int_{\Omega_{\alpha}} \delta \bfu^\mathrm{c} \cdot \rho_0 \:{\color{red}\ddot{\bfu}^{\mathrm{f}, \alpha}}  \: dV   + \int_{\Omega} \delta \bfu^\mathrm{c} \cdot \bfB \: dV + \int_{\Gamma^t} \delta \bfu^\mathrm{c} \cdot \Tilde{\bfT}  \: dA .  
\end{multlined} 
\end{equation}
The terms on the left-hand side of Eq.~\ref{eq:coarse-scale-weakform} correspond to kinetic energy and strain energy at the coarse scale, respectively, while the right-hand side terms correspond to the external work due to fine-scale dynamics, body force, and traction, respectively. Similarly, the fine-scale problem in each subdomain $\Omega_{\alpha}$ is obtained as:
\begin{equation} \label{eq:fine-scale-weakform}
\begin{multlined}
   \int_{\Omega_{\alpha}} \delta \bfu^{\mathrm{f},\alpha} \cdot \rho_0 \ddot{\bfu}^{\mathrm{f}, \alpha}\: dV \: + \int_{\Omega_{\alpha}} \boldsymbol{\nabla}_{\bfX} \delta \bfu^{\mathrm{f},\alpha} : \bfP \left(\bfX, t , {\color{red}\bfu^\mathrm{c}}, \bfu^{\mathrm{f},\alpha} \right) \: dV = \hspace{3.4cm} \\
   \quad - \int_{\Omega_{\alpha}} \delta \bfu^{\mathrm{f},\alpha} \cdot \rho_0 {\color{red} \ddot{\bfu}^\mathrm{c}}  \: dV 
   + \int_{\Omega_{\alpha}} \delta \bfu^{\mathrm{f},\alpha} \cdot \bfB \: dV + \int_{\Gamma^t_{\alpha}} \delta \bfu^{\mathrm{f},\alpha} \cdot \Tilde{\bfT}  \: dA.
\end{multlined}    
\end{equation}
The terms of Eq.~\ref{eq:fine-scale-weakform} are interpreted in an analogous fashion to the coarse scale problem. The traction term is only present at the subdomain boundaries that coincide with the exterior Neumann boundaries of the problem domain, as the fine-scale test function vanishes at all other subdomain boundaries (see Eq.~\eqref{fineScaleFunDef}). The coupling terms in the coarse and fine-scale problems are highlighted in red, shown respectively in Eqs.~\eqref{eq:coarse-scale-weakform} and \eqref{eq:fine-scale-weakform}. 

The boundary and initial conditions for the coarse-scale problem and the initial conditions for the fine-scale problems complete the multiscale governing equations. The initial state at the fine-scale is taken to be undeformed and stationary:
\begin{equation}
    \bfu^{\mathrm{f},\alpha} (\bfX, 0) = \mathbf{0}, \quad \dot{\bfu}^{\mathrm{f},\alpha} (\bfX, 0) = \mathbf{0}, \quad  \bfX \in \Omega_{\alpha} .
\end{equation}
The boundary and initial conditions for the coarse-scale problem are:
\begin{subequations}
\begin{gather}
    \bfu^\mathrm{c} (\bfX, t) = \tilde{\bfu} (\bfX, t), \quad \bfX \in \Gamma^u; \quad  \bfP (\bfX, t) \cdot \bfN = \Tilde{\bfT} (\bfX, t),  \quad \bfX \in \Gamma^t;  \\
    \bfu^\mathrm{c} (\bfX, 0) = \hat{\bfu} (\bfX), \quad \dot{\bfu}^\mathrm{c} (\bfX, 0) = \hat{\bfv} (\bfX), \quad  \bfX \in \Omega.  
\end{gather}    
\end{subequations}
The initial displacement $(\hat{\bfu})$ and velocity $(\hat{\bfv})$ fields are selected such that they can be accurately described by the coarse-scale discretization. 

%
\section{Computational Approximation of Two-Scale PDEs} \label{sec:computational_formulation_two_scale_pdes}
This section first describes the spatial and temporal discretization methods used for the multiscale governing equations in the proposed VME formulation. The details of the evaluation of non-standard element matrices and vectors due to multiscale discretization and estimation of stable time increments based on the time integration scheme are discussed later.

Numerical simulation of the dynamic response of complex microstructures, particularly in short-wavelength regimes, is often performed using higher-order elements or dissipative time integration schemes. In a related work, spectral elements up to the seventh order were employed for coarse-scale discretization, where each coarse-scale element corresponded to a unit cell~\cite{hu2020spectral}. In this study, we employ a strategy that resembles $h-$refinement by discretizing each unit cell with a patch of coarse-scale elements to accurately capture wave propagation in short-wavelength regimes.

\subsection{Spatial discretization} \label{sec:spatial_discretization}

Consider the decomposition of an enrichment subdomain, $\Omega_{\alpha}$ into a patch of coarse elements: $\Omega_{\alpha} = \bigcup_{E=1}^{n_\mathrm{ecp}} \Delta_{\alpha_E}^{\mathrm{c}}$, where a coarse finite element within that patch is denoted with $\Delta_{\alpha_E}^{\mathrm{c}}$. 
Using the classical Bubnov-Galerkin approach, the coarse-scale displacement, weighting function, and their gradients for a coarse-scale element are written as:
\begin{subequations} \label{eq:coarse_disp_and_grad_disp_matrices}
\begin{gather}
     \bfu^{\mathrm{c}}_{\alpha_E} = \bfN^{\mathrm{c}}_{\alpha_E} \bfd^{\mathrm{c}}_{\alpha_E} = \bfN^{\mathrm{c}}_{\alpha_E} \bfL^{\mathrm{c}}_{\alpha_E} \bfd^{\mathrm{c}}, \\
    [\bs{\nabla} \bfu^{\mathrm{c}}_{\alpha_E}] = \bfB^{\mathrm{c}}_{\alpha_E} \bfd^{\mathrm{c}}_{\alpha_E} = \bfB^{\mathrm{c}}_{\alpha_E} \bfL^{\mathrm{c}}_{\alpha_E} \bfd^{\mathrm{c}},
\end{gather}    
\end{subequations}
%
where $\bfu^{\mathrm{c}}_{\alpha_E} \left( \bfX, t \right) := \bfu^{\mathrm{c}} \left( \bfX \in \Delta^{\mathrm{c}}_{\alpha_E} , t \right)$. $[\cdot]$ denotes the vectorized form of the tensorial entities. Einstein's summation convention does not apply to the index, $\alpha_E$. $\bfN^{\mathrm{c}}_{\alpha_E}\left( \bfX \right)$ and $\bfB^{\mathrm{c}}_{\alpha_E} \left( \bfX \right)$ are the coarse-scale element shape function matrix and shape function gradient matrix, respectively, for the coarse-scale element, $\Delta^{\mathrm{c}}_{\alpha_E}$. $\bfd^{\mathrm{c}}_{\alpha_E}(t)$ is the nodal displacement vector for the coarse-scale element. The local vector is mapped to the corresponding global vector, $\bfd^{\mathrm{c}}(t)$ through the gather matrix, $\bfL^{\mathrm{c}}_{\alpha_E}$. The conventional forms typically used in the finite element literature is employed for these matrices (see e.g., Ref.~\cite{fish2007first}). The discretizations of the weight function and its gradient are similarly defined.

Each enrichment subdomain is discretized a second time, using $n_\mathrm{ef}$ fine-scale elements that resolves the underlying heterogeneous microstructure: $\Omega_{\alpha} = \bigcup_{e=1}^{n_\mathrm{ef}} \Delta_{\alpha_e}^{\mathrm{f}}$. The fine and coarse-scale discretizations are performed in a compatible fashion; i.e., $\Delta^{\mathrm{c}}_{\alpha_E} = \bigcup_{e\in I_{\alpha_E}} \Delta_{\alpha_e}^{\mathrm{f}}$ for any macroscale element. $I_{\alpha_E}$ denotes an index set of fine scale elements that resolve the coarse scale element, $\alpha_E$. The displacement, weighting function, and their gradients for a fine-scale element are given as:
\begin{subequations} \label{eq:fine_disp_and_grad_disp_matrices}
\begin{gather}
     \bfu^{\mathrm{f}}_{\alpha_e} = \bfN^{\mathrm{f}}_{\alpha_e} \bfd^{\mathrm{f}}_{\alpha_e} = \bfN^{\mathrm{f}}_{\alpha_e} \bfL^{\mathrm{f}}_{\alpha_e} \bfd^{\mathrm{f},{\alpha}}, \\
    [ \bs{\nabla} \bfu^{\mathrm{f}}_{\alpha_e} ] = \bfB^{\mathrm{f}}_{\alpha_e} \bfd^{\mathrm{f}}_{\alpha_e} = \bfB^{\mathrm{f}}_{\alpha_e} \bfL^{\mathrm{f}}_{\alpha_e} \bfd^{\mathrm{f},{\alpha}},
\end{gather}    
\end{subequations}
%
in which, $\bfu^{\mathrm{f}}_{\alpha_e}$, $\bfN^{\mathrm{f}}_{\alpha_e}$ and $\bfB^{\mathrm{f}}_{\alpha_e}$ are defined analogously to their coarse scale counterparts. The gather matrix, $\bfL^{\mathrm{f}}_{\alpha_e}$ maps the nodal displacement vector associated with element, $\alpha_e$ (i.e., $\bfd^{\mathrm{f}}_{\alpha_e}$) with the nodal displacement vector for the whole enrichment subdomain, $\bfd^{\mathrm{f},{\alpha}}$.

Using Eqs.~\eqref{eq:coarse_disp_and_grad_disp_matrices} and \eqref{eq:fine_disp_and_grad_disp_matrices}, the discretized forms of Eqs.~\eqref{eq:coarse-scale-weakform} and \eqref{eq:fine-scale-weakform} are obtained as:
\begin{subequations} \label{eq:ode_coarse_fine_both}
\begin{gather}
    \bfM^{\mathrm{c}} \ddot{\bfd}^\mathrm{c} + \sum_{\alpha =1}^{n_{\mathrm{es}}} \bfM^{\mathrm{c} \mathrm{f}_{\alpha}} \ddot{\bfd}^{\mathrm{f}_\alpha} +  \bff^{\mathrm{c}}_{\mathrm{int}} \left( \bfd^\mathrm{c}, \{ \bfd^{\mathrm{f},\alpha} \} \right) = \bff^\mathrm{c}_{\mathrm{ext}}; \label{eq:ode_coarse_scale} \\ 
    \bfM^{\mathrm{f}_{\alpha}} \ddot{\bfd}^{\mathrm{f}_{\alpha}} + \bfM^{\mathrm{f}_{\alpha} \mathrm{c}} \ddot{\bfd}^{\mathrm{c}, \alpha} + \bff^{\mathrm{f}_\alpha}_{\mathrm{int}} \left( \bfd^{\mathrm{c},\alpha}, \bfd^{\mathrm{f}, {\alpha}} \right) = \bff^{\mathrm{f}_{\alpha}}_{\mathrm{ext}}, \quad \; \alpha = 1 \; \text{to} \; n_{\mathrm{es}};\label{eq:ode_fine_scale}
\end{gather}    
\end{subequations}
where $\bfd^{\mathrm{c, \alpha}}$ denotes the nodal displacement vector corresponding to the enrichment subdomain, $\Omega_{\alpha}$ and related to the total displacement vector through a gather matrix: $\bfd^{\mathrm{c, \alpha}} = \mathbf{L}^{\mathrm{c}}_\alpha \bfd^{\mathrm{c}}$. $\bfM^{\mathrm{c}}$ and $\bfM^{\mathrm{f}_{\alpha}}$ are the coarse- and fine-scale mass matrices, respectively; $\bfM^{\mathrm{c} \mathrm{f}_{\alpha}}$ and $\bfM^{\mathrm{f}_{\alpha} \mathrm{c}}$ are the mass matrices that describe scale interactions. The internal force vectors at the coarse and fine-scales are given by $\bff^{c}_{\mathrm{int}}$, $\bff^{\mathrm{f}_\alpha}_{\mathrm{int}}$ respectively, and the external force vectors at coarse and fine-scale are $\bff^c_{\mathrm{ext}}$ and $\bff^{\mathrm{f}_{\alpha}}_{\mathrm{ext}}$, respectively. These are obtained by assembling the element-level matrices and internal and external force vectors as shown below:
\begin{subequations}
\begin{align}
    & \bfM^{\mathrm{c}} = \sum_{\alpha,E} \left( \bfL^{\mathrm{c}}_{\alpha_E}\right)^T  \bfM^{\mathrm{c}}_{\alpha_E}  \bfL^{\mathrm{c}}_{\alpha_E}; \quad 
    \bfM^{\mathrm{c} \mathrm{f}_{\alpha}} = \left( \bfM^{ \mathrm{f}_{\alpha} \mathrm{c}}\right)^T = \sum_{E, \: e\in I_{\alpha_E}} \left( \bfL^{\mathrm{c}}_{\alpha_E}\right)^T \bfM_{E, e}^{\alpha} \bfL^\mathrm{f}_{\alpha_e}; \\
    & \bfM^{\mathrm{f}_\alpha} = \sum_{e} \left( \bfL^\mathrm{f}_{\alpha_e} \right)^T \bfM_{e}^{\mathrm{f}_\alpha} \bfL^\mathrm{f}_{\alpha_e}; \quad 
    \bff^{\mathrm{c}}_{\mathrm{int}} = \sum_{\alpha,E} \left( \bfL^{\mathrm{c}}_{\alpha_E}\right)^T \bff^{\mathrm{c}, \alpha_E}_{\mathrm{int}}; \\  
    & \bff^{\mathrm{f}_\alpha}_{\mathrm{int}} = \sum_{e} \left( \bfL^\mathrm{f}_{\alpha_e} \right)^T \bff^{\mathrm{f}_\alpha, e}_{\mathrm{int}}; \quad 
     \bff^{\mathrm{c}}_{\mathrm{ext}} = \sum_{\alpha,E} \left( \bfL^{\mathrm{c}}_{\alpha_E} \right)^T \bff^{ \mathrm{c}, \alpha_E}_{\mathrm{ext}}; \quad 
     \bff^{\mathrm{f}_{\alpha}}_{\mathrm{ext}} = \sum_{e} \left( \bfL^\mathrm{f}_{\alpha_e} \right)^T \bff^{\mathrm{f}_\alpha, e}_{\mathrm{ext}}.
\end{align}
\end{subequations}
The corresponding coarse-scale element-level matrices and force vectors are obtained as:
\begin{subequations} \label{eq:element-level-matrices-vectors-coarse}
\begin{align}
    & \bfM^{\mathrm{c}}_{\alpha_E} = \int_{\Omega_{\alpha_E}} \left( \bfN^{\mathrm{c}}_{\alpha_E} \right)^T \rho_0  \bfN^{\mathrm{c}}_{\alpha_E} \: dV; \quad \bff^{\mathrm{c},\alpha_E}_{\mathrm{int}} = \int_{\Omega_{\alpha_E}} \left( \bfB^{\mathrm{c}}_{\alpha_E} \right)^T  [\bfP] \: dV; \\ 
    & \bff^{\mathrm{c}, \alpha_E}_{\mathrm{ext}} = \int_{\Omega_{\alpha_E}} \left( \bfN^{\mathrm{c}}_{\alpha_E} \right)^T \bfB \: dV + \int_{\Gamma^t}  \left( \bfN^{\mathrm{c}}_{\alpha_E} \right)^T \tilde{\bfT} \: dA \: ;
\end{align}    
\end{subequations}
where $[\bfP]$ is the vectorized form of the first Piola-Kirchhoff stress tensor. The vectorized form (distinct from the Voigt notation) includes all components of the stress tensor since it is not symmetric. In the case of a composite microstructure where the density spatially varies, the integrations for $\bfM^{\mathrm{c}}_{\alpha_E}$ and $\bff^{\mathrm{c}, \alpha_E}_{\mathrm{ext}}$ are performed by further discretizing the coarse-scale elements into its fine-scale counterparts or by approximating the integration by averaging the density. The integration for the internal force vector is performed over the underlying fine-scale grid to capture the stress variations within the microstructure.

The fine-scale element-level matrices and force vectors are obtained as:
\begin{subequations} \label{eq:element-level-matrices-vectors-fine}
\begin{align}
    & \bfM_{e}^{\mathrm{f}_\alpha} = \int_{\Omega_{\alpha_e}} \left( \bfN^\mathrm{f}_{\alpha_e} \right)^T \rho_0 \bfN^\mathrm{f}_{\alpha_e} \: dV; \quad  \bfM_{E, e}^{\alpha} = \int_{\Omega_{\alpha_e}} \left( \bfN^{\mathrm{c}}_{\alpha_E} \right)^T \rho_0 \bfN^\mathrm{f}_{\alpha_e}  \: dV; \\ 
    & \bff^{\mathrm{f}_\alpha, e}_{\mathrm{int}} = \int_{\Omega_{\alpha_e}} \: \left( \bfB^\mathrm{f}_{\alpha_e} \right)^T [\bfP] \: dV; \quad  \bff^{\mathrm{f}_\alpha, e}_{\mathrm{ext}} = \int_{\Omega_{\alpha_e}} \left( \bfN^\mathrm{f}_{\alpha_e} \right)^T \bfB \: dV + \int_{\Gamma^t_{\alpha_e}}  \left( \bfN^\mathrm{f}_{\alpha_e} \right)^T \tilde{\bfT} \: dA .
\end{align}    
\end{subequations}
%
%
The evaluation of $\bfM^{\mathrm{c}}_{\alpha_E}$ and $\bfM_{E, e}^{\alpha}$, $\bff^{\mathrm{c}, \alpha_E}_{\mathrm{int}}$ and $\bff^{\mathrm{f}_\alpha, e}_{\mathrm{int}}$ is non-standard and further discussed in Section~\ref{sec:element_matrices_and_vectors}. The remaining entities in Eqs.~\eqref{eq:element-level-matrices-vectors-coarse} and~\eqref{eq:element-level-matrices-vectors-fine} are evaluated using the standard element-level integration procedure.
%
%
\subsection{Time integration}
The coupled two-scale semi-discrete multiscale equations given in Eq.~\eqref{eq:ode_coarse_fine_both} are integrated iteratively using an operator-split procedure until convergence is obtained for each time step. Within the operator-split procedure, various time-integration schemes for the coarse-scale and fine-scale equations are employed in this work, namely: (1) explicit update for both equations using the central difference method, (2) explicit \citet{noh2013explicit} scheme for both equations, and (3) mixed explicit-implicit time integration scheme with explicit integration for the coarse-scale equations using \citet{noh2013explicit} scheme and implicit integration for the fine-scale equations using \citet{bathe2005composite} scheme. The performance of these integration schemes is assessed in Section~\ref{sec:cost_comparison_integration_schemes}.

The following notation is adopted: Consider that the global displacement, velocity, and acceleration vectors at the coarse and fine scales are known at time, $t_n$ (also referred to as the $n^{\mathrm{th}}$ time step). These vectors are denoted by $\big(\bfd^\mathrm{c}_n, \dot{\bfd}^\mathrm{c}_n, \ddot{\bfd}^\mathrm{c}_n \big)$ and $\big( {\bfd^{\mathrm{f}, \alpha}_n}, {\dot{\bfd}^{\mathrm{f}, \alpha}_n}, {\ddot{\bfd}^{\mathrm{f},\alpha}_n} \big)$ respectively. The time integration results in the corresponding coarse and fine-scale fields at the $(n+1)^{\mathrm{th}}$ time step, which are denoted as $\big(\bfd^\mathrm{c}_{n+1}, \dot{\bfd}^\mathrm{c}_{n+1}, \ddot{\bfd}^\mathrm{c}_{n+1} \big)$ and $\big( {\bfd^{\mathrm{f}, \alpha}_{n+1}}, {\dot{\bfd}^{\mathrm{f}, \alpha}_{n+1}}, {\ddot{\bfd}^{\mathrm{f},\alpha}_{n+1}} \big)$ respectively. The state of a vector at the ${k}^{\mathrm{th}}$ iteration of the operator-split procedure within [$t_n,t_{n+1}$] is denoted by a second subscript \big(e.g., $\ddot{\bfd}^\mathrm{c}_{n+1,k}$\big). 
%
%
\subsubsection{Explicit-explicit central difference method (EE-CDM)} \label{sec:Explicit-central-difference-method}
In this method, both the coarse- and the fine-scale equations (Eqs.~\eqref{eq:ode_coarse_fine_both}a-b) are integrated using the explicit central difference method. The implementation procedure is provided in Algorithm~\ref{alg:EE-CDM}. At a given time increment, $t_n$, the algorithm updates the multiscale nodal displacement vectors, first. The acceleration vectors are computed iteratively until convergence. Convergence is assessed based on discrete $L^{\infty}$ norm, in order to ensure that the accuracy tolerance is strictly enforced component-wise. Upon convergence, the velocity vectors are updated. EE-CDM is the most straightforward and efficient method among the three investigated, but, similar to the single-scale implementation, the multiscale implementation proposed here also could suffer from instability. 

\begin{algorithm} [htb!]
 \small
    \caption{Implementation of EE-CDM.}
    \label{alg:EE-CDM}
    \begin{algorithmic}[1]
        \State \textbf{Input:} Given the state at $t_n$ and the external force vectors at $t_{n+1}$. 
        \State Estimate a stable time increment, $\Delta t$, for the multiscale problem as described in Section~\ref{sec:stable_time_increment_multiscale_problem}. Set $t_{n+1}=t_n+\Delta t$ and iteration count $k = 1$.
        \State Set the initial guess for the fine-scale acceleration: $\ddot{\bfd}^{\mathrm{f}, \alpha}_{n+1, 0}=\ddot{\bfd}^{\mathrm{f}, \alpha}_{n}$.
        \State Update the displacement fields:
\begin{align*}
    & \bfd^{\mathrm{f}, \alpha}_{n+1} = \bfd^{\mathrm{f}, \alpha}_{n} + \Delta t \: \dot{\bfd}^{\mathrm{f}, \alpha}_n + \frac{(\Delta t)^2}{2} \: \ddot{\bfd}^{\mathrm{f}, \alpha}_n; \hspace{0.3cm} \alpha=1,2,\dots,n_{\mathrm{es}},  \\
    & \bfd^\mathrm{c}_{n+1} = \bfd^\mathrm{c}_{n} + \Delta t \: \dot{\bfd}^\mathrm{c}_n + \frac{(\Delta t)^2}{2} \: \ddot{\bfd}^\mathrm{c}_n.
\end{align*}
    \State At any arbitrary iteration count, $k$: \label{cdm_step}
    \begin{algsubstates}
        \State Explicitly integrate the coarse-scale acceleration: \label{cdm-coarse-scale}
\begin{equation*}
    \bfM^{\mathrm{c}} \ddot{\bfd}^\mathrm{c}_{n+1,k} + \sum_{\alpha=1}^{n_{\mathrm{ec}}} \bfM^{\mathrm{c} \mathrm{f}_{\alpha}} \ddot{\bfd}^{\mathrm{f},\alpha}_{n+1,k-1} + \bff^{\mathrm{c}}_{\mathrm{int}} \left( \mathbf{d}^\mathrm{c}_{n+1} , \bfd^{\mathrm{f}}_{n+1} \right) = \bff^c_{\mathrm{ext}}.
\end{equation*}    
        \State Explicitly integrate the fine-scale acceleration for each subdomain, $\alpha$: \label{cdm-fine-scale}
\begin{equation*}
    \bfM^{\mathrm{f}_{\alpha}} \ddot{\bfd}^{\mathrm{f}, \alpha}_{n+1, k} + \bfM^{\mathrm{f}_{\alpha} \mathrm{c}} \mathbf{L}^{\mathrm{c}}_\alpha \ddot{\bfd}^{\mathrm{c}}_{n+1,k} +  \bff^{\mathrm{f}_\alpha}_{\mathrm{int}} \left( \mathbf{d}^{\mathrm{c}, \alpha}_{n+1} , 
 \bfd^{\mathrm{f},\alpha}_{n+1} \right)  = \bff^{\mathrm{f}_{\alpha}}_{\mathrm{ext}}.
\end{equation*}
    \State Check for convergence \label{state-error-cdm}:
\begin{equation*}
    e^{\mathrm{f},\alpha}_{n+1}:=\frac{||\ddot{\bfd}^{\mathrm{f},\alpha}_{n+1,k} - \ddot{\bfd}^{\mathrm{f},\alpha}_{n+1,k-1}||_{\infty}}{||\ddot{\bfd}^{\mathrm{f},\alpha}_{n+1,k-1}||_{\infty}}  < \mathrm{tol}_\mathrm{f} \:; \quad 
     e^{\mathrm{c}}_{n+1}:= \frac{||\ddot{\bfd}^{\mathrm{c}}_{n+1,k} - \ddot{\bfd}^{\mathrm{c}}_{n+1,k-1}||_{\infty}}{||\ddot{\bfd}^{\mathrm{c}}_{n+1,k-1}||_{\infty}}  < \mathrm{tol}_\mathrm{c} \: .\label{error-cdm}
\end{equation*}
\end{algsubstates}
        \State \textbf{If} {\ref{state-error-cdm} is met} \textbf{then} $\ddot{\bfd}^{\mathrm{f},\alpha}_{n+1} \gets \ddot{\bfd}^{\mathrm{f},\alpha}_{n+1,k}$ and $\ddot{\bfd}^{\mathrm{c}}_{n+1} \gets \ddot{\bfd}^{\mathrm{c}}_{n+1,k}$.
        \State \textbf{Else} $k \gets k+1$, and \textbf{go to} \ref{cdm_step} for the next iteration.
        \State Update the velocity fields:
\begin{align*}
    & \dot{\bfd}^{\mathrm{f}, \alpha}_{n+1} = \dot{\bfd}^{\mathrm{f}, \alpha}_n +  \frac{\Delta t}{2} \left( \ddot{\bfd}^{\mathrm{f},\alpha}_n + \ddot{\bfd}^{\mathrm{f},\alpha}_{n+1} \right); \hspace{0.3cm} \alpha=1,2,\dots,n_{\mathrm{es}}, \\
    & \dot{\bfd}^\mathrm{c}_{n+1} = \dot{\bfd}^\mathrm{c}_n +  \frac{\Delta t}{2} \left( \ddot{\bfd}^\mathrm{c}_n + \ddot{\bfd}^\mathrm{c}_{n+1} \right).
\end{align*}
        \State \Return $\big(\bfd^\mathrm{c}_{n+1}, \dot{\bfd}^\mathrm{c}_{n+1}, \ddot{\bfd}^\mathrm{c}_{n+1} \big)$ and $\big( {\bfd^{\mathrm{f}, \alpha}_{n+1}}, {\dot{\bfd}^{\mathrm{f}, \alpha}_{n+1}}, {\ddot{\bfd}^{\mathrm{f},\alpha}_{n+1}} \big)$.
\end{algorithmic}
\end{algorithm}


%
\subsubsection{Explicit-explicit sub-step method (EE-SSM)} \label{sec:Explicit-sub-step-method}

Despite its large time step stability limit among explicit schemes, it is well-known that dispersion errors in high-frequency modes can lead to degraded solution accuracy when the central difference method is employed. The integration procedure for EE-SSM, which is used to alleviate spurious high frequency dispersion, is described in Algorithm~\ref{alg:EE-SSM}. The coarse and fine-scale problems are integrated using an explicit sub-step time integration scheme proposed by \citet{noh2013explicit}. In this scheme, the coarse and fine-scale fields are integrated in two steps, first at a sub-step denoted by $t_{n+p}$ ($0<p<1$) and then at the full step denoted by $t_{n+1}$. In both updates, the operator-split procedure is used to iteratively solve for coarse- and fine-scale fields. 
%
\begin{algorithm} [htb!]
    \small
    \caption{Implementation of EE-SSM.}
    \label{alg:EE-SSM}
    \begin{algorithmic}[1]
        \State \textbf{Input:} Given the state at $t_n$ and the external force vectors at $t_{n+1}$. 
        \State Estimate a stable time increment, $\Delta t$, and set the sub-step ratio $p$. $t_{n+p}=t_n+p \Delta t$ and iteration count $k = 1$. Set the constants of integration:
\begin{equation*}
    \begin{aligned}
    & q_1 = \frac{1 -2p}{2p(1-p)}; \quad q_2 = \frac{1}{2} - p q_1; \quad q_0 = -q_1 - q_2 + \frac{1}{2}; \quad a_0 = p \Delta t; \quad a_1 = \frac{1}{2} \left( p \Delta t \right)^2; \\
     &   a_2 = {a_0}/{2}; \quad a_3 = (1-p)\Delta t; \quad a_4 = \frac{1}{2} a_3^2; \quad a_5 = q_0 a_3; \quad  a_6 = \left( 0.5 + q_1 \right) a_3; \quad a_7 = q_2 a_3.
    \end{aligned}
\end{equation*}
\noindent \textbf{Sub-step}: 
\State Compute the external force vectors at $t_{n+p}$.
\State Set the initial guess for the fine-scale acceleration: $\ddot{\bfd}^{\mathrm{f}, \alpha}_{n+p, 0}=\ddot{\bfd}^{\mathrm{f}, \alpha}_{n}$.
\State Update the fine- and coarse-scale displacement fields: \label{state_eessm_disp_substep}
\begin{equation*}
    \bfd^{\mathrm{f},\alpha}_{n+p} = \bfd^{\mathrm{f},\alpha}_{n} + a_0 \: \dot{\bfd}^{\mathrm{f}, \alpha}_n + a_1 \ddot{\bfd}^{\mathrm{f},\alpha}_n; \hspace{0.3cm} \bfd^{\mathrm{c}}_{n+p} = \bfd^{\mathrm{c}}_{n} + a_0 \dot{\bfd}^{\mathrm{c}}_n + a_1 \ddot{\bfd}^{\mathrm{c}}_n.
\end{equation*}
\State At an arbitrary increment, $k>0$: \label{explicit_explicit_substep}
\begin{algsubstates}
\State Explicitly integrate the coarse- and fine-scale accelerations: \label{state-eesm-substep-coarse-fine-scale-accel}
\begin{gather*}
    \bfM^{\mathrm{c}} \ddot{\bfd}^\mathrm{c}_{n+p,k} + \sum_{\alpha=1}^{n_{\mathrm{es}}} \bfM^{\mathrm{c} \mathrm{f}_{\alpha}} \ddot{\bfd}^{\mathrm{f},\alpha}_{n+p,k-1} + \bff^{\mathrm{c}}_{\mathrm{int}} \left( \mathbf{d}^\mathrm{c}_{n+p} , \bfd^{\mathrm{f}}_{n+p} \right) = \bff^\mathrm{c}_{\mathrm{ext}}, \\
    \bfM^{\mathrm{f}_{\alpha}} \ddot{\bfd}^{\mathrm{f}, \alpha}_{n+p, k} + \bfM^{\mathrm{f}_{\alpha} \mathrm{c}} \mathbf{L}^{\mathrm{c}}_\alpha \ddot{\bfd}^{\mathrm{c}}_{n+p,k} +  \bff^{\mathrm{f}_\alpha}_{\mathrm{int}} \left( \mathbf{d}^{\mathrm{c}, \alpha}_{n+p} , 
 \bfd^{\mathrm{f},\alpha}_{n+p} \right)  = \bff^{\mathrm{f}_{\alpha}}_{\mathrm{ext}}.
\end{gather*}    

\State Check for convergence: \label{state-error-substep-explicit-explicit}
\begin{equation*}
    e^{\mathrm{f},\alpha}_{n+p}  < \mathrm{tol}_\mathrm{f} \:; \quad 
     e^{\mathrm{c}}_{n+p}  < \mathrm{tol}_\mathrm{c} \: .\label{error-cdm_substep}
\end{equation*}
   
\end{algsubstates}
\State \textbf{If} {\ref{state-error-substep-explicit-explicit} is met} \textbf{then}
            $\hspace{0.2cm} \ddot{\bfd}^{\mathrm{f},\alpha}_{n+p} \gets \ddot{\bfd}^{\mathrm{f},\alpha}_{n+p,k}$ and $\ddot{\bfd}^{\mathrm{c}}_{n+p} \gets \ddot{\bfd}^{\mathrm{c}}_{n+p,k}$.
\State \textbf{Else}
            $\hspace{0.2cm} k \gets k+1$, and \textbf{go to} \ref{explicit_explicit_substep} for the next iteration. 
\State Update the velocity fields: \label{state-eesm-velocity-substep}
\begin{equation*}
    \dot{\bfd}^{\mathrm{f}, \alpha}_{n+p} = \dot{\bfd}^{\mathrm{f}, \alpha}_n +  a_2 \big( \ddot{\bfd}^{\mathrm{f},\alpha}_n + \ddot{\bfd}^{\mathrm{f},\alpha}_{n+p} \big); \hspace{0.3cm} \dot{\bfd}^\mathrm{c}_{n+p} = \dot{\bfd}^\mathrm{c}_n +  a_2 \big( \ddot{\bfd}^\mathrm{c}_n + \ddot{\bfd}^\mathrm{c}_{n+p} \big).
\end{equation*}

    \noindent \textbf{Full-step}: 
    \State Set $k=1$; Compute the external force vectors at $t_{n+1}$ and set the initial guess for the fine-scale acceleration: $\ddot{\bfd}^{\mathrm{f}, \alpha}_{n+1, 0}=\ddot{\bfd}^{\mathrm{f}, \alpha}_{n+p}$.
    \State Update the fine- and coarse-scale displacement fields: \label{state-eesm-displacement-fullstep}
\begin{equation*}
    \bfd^{\mathrm{f},\alpha}_{n+1} = \bfd^{\mathrm{f},\alpha}_{n+p} + a_3 \dot{\bfd}^{\mathrm{f}, \alpha}_{n+p} + a_4 \ddot{\bfd}^{\mathrm{f},\alpha}_{n+p}; \hspace{0.3cm} \bfd^{\mathrm{c}}_{n+1} = \bfd^{\mathrm{c}}_{n+p} + a_3 \dot{\bfd}^{\mathrm{c}}_{n+p} + a_4 \ddot{\bfd}^{\mathrm{c}}_{n+p}.
\end{equation*}
    \State Iterate the coarse- and fine-scale accelerations until convergence by following step~\ref{cdm_step} of the Algorithm \ref{alg:EE-CDM} provided for the central difference method.  
    \State Update the velocity fields: \label{state-eesm-velocity-fullstep}
\begin{equation*}
    \dot{\bfd}^{\mathrm{f}, \alpha}_{n+1} = \dot{\bfd}^{\mathrm{f}, \alpha}_{n+p} + a_5 \ddot{\bfd}^{\mathrm{f}, \alpha}_n +  a_6 \ddot{\bfd}^{\mathrm{f},\alpha}_{n+p} + a_7 \ddot{\bfd}^{\mathrm{f},\alpha}_{n+1}; \hspace{0.3cm} \dot{\bfd}^\mathrm{c}_{n+1} = \dot{\bfd}^\mathrm{c}_{n+p} + a_5 \ddot{\bfd}^\mathrm{c}_{n} + a_6  \ddot{\bfd}^\mathrm{c}_{n+p} + a_7 \ddot{\bfd}^\mathrm{c}_{n+1}.
\end{equation*}
\State \Return $\big(\bfd^\mathrm{c}_{n+1}, \dot{\bfd}^\mathrm{c}_{n+1}, \ddot{\bfd}^\mathrm{c}_{n+1} \big)$ and $\big( {\bfd^{\mathrm{f}, \alpha}_{n+1}}, {\dot{\bfd}^{\mathrm{f}, \alpha}_{n+1}}, {\ddot{\bfd}^{\mathrm{f},\alpha}_{n+1}} \big)$.
    
\end{algorithmic}
\end{algorithm}

\subsubsection{Explicit-implicit sub-step method (EI-SSM)} \label{sec:Implicit-sub-step-method}

We also explore an explicit-implicit sub-step time integration method, where the coarse-scale problem is integrated explicitly using the Noh and Bathe~\citep{noh2013explicit} scheme and the fine-scale problems are implicitly integrated using the Bathe and Baig~\citep{bathe2005composite} scheme. 
The resulting multiscale scheme is not unconditionally stable as the coarse-scale equation is updated explicitly, but the stable time increment of the coarse-scale problem is expected to be larger than that of fine-scale problems. Hence, the time increment for this scheme is expected to be governed by accuracy requirements rather than the stability constraints of the fine-scale equations. This approach could be advantageous compared to the explicit-explicit sub-step method, if the time step required to ensure accuracy is sufficiently larger than the stability limit such that the overall integration method is computationally beneficial despite the added computational cost of the implicit update. A microstructure with high stiffness contrast, where the stability of the fine-scale explicit update severely constrains the time step size, is such a problem. 

As the model exhibits geometric and material nonlinearity, the Newton-Raphson method is employed to solve the nonlinear fine-scale equations in EI-SSM. To solve the fine-scale equations using the Newton-Raphson method, the first variation of the residual given in Eq.~\eqref{eq:fine-scale-weakform} is taken along the direction $\big( \tilde{d}\bfu^{\mathrm{f},\alpha}, \tilde{d}\ddot{\bfu}^{\mathrm{f},\alpha} \big)$ to obtain the corresponding Jacobian. The incremental fine-scale fields are discretized analogously as described  in Eq.~\eqref{eq:fine_disp_and_grad_disp_matrices} to obtain $\big( \tilde{d}\bfd^{\mathrm{f},\alpha}, \tilde{d}\ddot{\bfd}^{\mathrm{f},\alpha} \big)$ and using the classical Bubnov-Galerkin approach, the semi-discrete version of the Jacobian is obtained. The semi-discrete Jacobian and the semi-discrete residual (given in Eq.~\eqref{eq:ode_coarse_fine_both}b) form the linearized system of equations for the incremental fine-scale fields as shown below:
\begin{equation} \label{eq:discretized_linear_weak_form}
 \bfM^{\mathrm{f}_\alpha } \tilde{d} \ddot{\bfd}^{\mathrm{f}, \alpha}  + \bfK^{\mathrm{f}_\alpha} \big( \bfd^{\mathrm{c}, \alpha} , \bfd^{\mathrm{f}, \alpha} \big) \: \tilde{d} \bfd^{\mathrm{f}, \alpha}  = - \left( \bfM^{\mathrm{f}_\alpha} \ddot{\bfd}^{\mathrm{f}, \alpha}  + \bff^{\mathrm{f}_\alpha}_{\mathrm{int}} \big( \bfd^{\mathrm{c},\alpha}, \bfd^{\mathrm{f},\alpha} \big) + \bfM^{\mathrm{f}_{\alpha} \mathrm{c}} \ddot{\bfd}^{\mathrm{c}}  - \bff^{\mathrm{f}_{\alpha}}_{\mathrm{ext}} \right), 
\end{equation}
where $\bfK^{\mathrm{f}_\alpha}$ is the fine-scale tangent stiffness matrix. It can be obtained using the corresponding element-level matrix as shown below:
\begin{equation}
\begin{aligned}
    & \bfK^{\mathrm{f}_\alpha} = \sum_{e} \left( \bfL^\mathrm{f}_{\alpha_e} \right)^T \bfK_e^{\mathrm{f}_\alpha} \bfL^\mathrm{f}_{\alpha_e}, \quad \text{where} 
    & \bfK_e^{\mathrm{f}_\alpha} = \int_{\Omega_{\alpha_e}} \left( \bfB^\mathrm{f}_{\alpha_e} \right)^T  [\bfD] \: \bfB^\mathrm{f}_{\alpha_e} dV.
\end{aligned}
\end{equation}
The $[\bfD]$ is the matrix form of the fourth-order tensor ${\partial \bfP}/{\partial \bfF} \left( \bfd^{\mathrm{c}, \alpha}, \bfd^{\mathrm{f}, \alpha} \right)$. The fully-discrete version of the linearized system of fine-scale fields is obtained by substituting the incremental acceleration in terms of incremental displacement in Eq.~\eqref{eq:discretized_linear_weak_form}. To do the same, the implicit acceleration update is linearized in terms of the unknown displacement field, both at the sub-step and full-step. For the sub-step, the following linear system of equations is obtained which is solved iteratively using the Newton-Raphson (N-R) method:
\begin{equation} \label{eq:implicit-sub-step-newton-raphson}
\begin{aligned}
       &  \left( \frac{4}{p^2 \Delta t^2} \bfM^{\mathrm{f}_\alpha}  + \bfK^{\mathrm{f}_\alpha } \left( \bfd^{\mathrm{c},\alpha}_{n+p}, \big( \bfd^{\mathrm{f},\alpha}_{n+p,k} \big)^{(i-1)} \right) \right) \big( {\tilde{d} \bfd}^{\mathrm{f}, \alpha}_{n+p,k}\big)^{(i)}  =
        \\  & \: - \bfM^{\mathrm{f}_\alpha} \big(\ddot{\bfd}^{\mathrm{f}, \alpha}_{n+p,k}\big)^{(i-1)}   - \bff^{\mathrm{f}_\alpha}_{\mathrm{int}} \left( \bfd^{\mathrm{c},\alpha}_{n+p}, \big( \bfd^{\mathrm{f},\alpha}_{n+p,k} \big)^{(i-1)} \right) - \bfM^{\mathrm{f}_{\alpha} \mathrm{c}} \ddot{\bfd}^{\mathrm{c}}_{n+p,k} + \bff^{\mathrm{f}_{\alpha}}_{\mathrm{ext}}, 
\end{aligned}
\end{equation}
where $(i)$ superscript corresponds to the vectorial fields at the $i^\mathrm{th}$ iteration of the N-R method. Similarly, the following linear system of equations is obtained at the full-step:
\begin{equation}\label{eq:implicit-full-step-newton-raphson}
    \begin{aligned}
        & \left( c_3 c_3 \bfM^{\mathrm{f}_\alpha}  +  \bfK^{\mathrm{f}_\alpha} \left( \bfd^{c,\alpha}_{n+1}, \big( \bfd^{\mathrm{f},\alpha}_{n+1,k} \big)^{(i-1)} \right) \right) \big( {\tilde{d} \bfd}^{\mathrm{f}, \alpha}_{n+1,k}\big)^{(i)} =
        \\ & \: - \bfM^{\mathrm{f}_\alpha} \big(\ddot{\bfd}^{\mathrm{f}, \alpha}_{n+1,k}\big)^{(i-1)}  
         - \bff^{\mathrm{f}_\alpha}_{\mathrm{int}} \left( \bfd^{\mathrm{c},\alpha}_{n+1}, \big( \bfd^{\mathrm{f},\alpha}_{n+1,k} \big)^{(i-1)} \right) - \bfM^{\mathrm{f}_{\alpha} \mathrm{c}} \ddot{\bfd}^{\mathrm{c}}_{n+1,k} + \bff^{\mathrm{f}_{\alpha}}_{\mathrm{ext}},
    \end{aligned}
    \end{equation}
which can be similarly solved using N-R method.

The implicit update of the fine-scale problem for the sub-step and full-step are described in Algorithm \ref{alg:EI-SSM}. The coarse-scale problem update remains the same as discussed in Algorithm \ref{alg:EE-SSM}. The sub-step ratio, $p$, is taken to be the same for both coarse and fine-scale problems for consistent evaluation of fields for both scales at the sub-step and full-step. Unlike EE-CDM and EE-SSM, the stable time increment is chosen based on the coarse-scale problem. 
\begin{algorithm} [htb!]
\small
\caption{Implementation of EI-SSM.}
\label{alg:EI-SSM}
\begin{algorithmic}[1]
\State \textbf{Input:} Given the state at $t_n$ and the external force vectors at $t_{n+1}$. 
\State Estimate a stable time increment, $\Delta t$, and set the sub-step ratio $p$. $t_{n+p}=t_n+p \Delta t$ and iteration count $k = 1$. Get the integration constants for the implicit scheme as given below:
\begin{equation*}
    \begin{aligned}
      & c_1 = ({1-p})/({p \: \Delta t});  \quad c_2 = {-1}/((1-p) p \: \Delta t); \quad c_3 = (2-p)/((1-p) \Delta t).
    \end{aligned}
\end{equation*}
\noindent \textbf{Sub-step}: 
\State Compute the external force vector at $t_{n+p}$; and set the initial guesses: $\bfd^{\mathrm{f}, \alpha}_{n+p, 0}=\bfd^{\mathrm{f}, \alpha}_{n}$, $\ddot{\bfd}^{\mathrm{f},\alpha}_{n+p, 0} = \ddot{\bfd}^{\mathrm{f},\alpha}_{n}$.
\State At an arbitrary increment, $k>0$: \label{state-explicit-implicit-substep}
\begin{algsubstates}
    \State The coarse-scale acceleration vector is updated as given in step \ref{state_eessm_disp_substep} of Algorithm \ref{alg:EE-SSM}.
    \State Solve Eq.\eqref{eq:implicit-sub-step-newton-raphson} iteratively for the fine-scale displacement and acceleration corresponding to each subdomain, $\alpha$, and the updates for the $i^\mathrm{th}$ N-R iteration are shown below:
    \begin{align*}
        & \big( {\bfd}^{\mathrm{f}, \alpha}_{n+p,k} \big)^{(i)} = \big( {\bfd}^{\mathrm{f}, \alpha}_{n+p,k} \big)^{(i-1)} + \big( {\tilde{d} \bfd}^{\mathrm{f}, \alpha}_{n+p,k}\big)^{(i)}; \\
        & \big( \ddot{\bfd}^{\mathrm{f}, \alpha}_{n+p,k} \big)^{(i)} = \left( \big( \bfd^{\mathrm{f}, \alpha}_{n+p,k} \big)^{(i)}  - \bfd^{\mathrm{f}, \alpha}_{n} - \dot{\bfd}^{\mathrm{f}, \alpha}_{n} p \Delta t \right) \frac{4}{p^2 \Delta t^2} - \ddot{\bfd}^{\mathrm{f}, \alpha}_{n}.
    \end{align*}
    The Newton iterations are performed until the norm of the discrete residual vector (the right-hand side of the Eq.~\eqref{eq:implicit-sub-step-newton-raphson}) is below a set tolerance value.
    \State Check the following error quantities in addition to those mentioned in step \ref{state-error-substep-explicit-explicit} of Algorithm \ref{alg:EE-SSM}: \label{state-error-substep-explicit-implicit}
    \begin{equation*}
        \frac{||{\bfd}^{\mathrm{f},\alpha}_{n+p,k} - {\bfd}^{\mathrm{f},\alpha}_{n+p,k-1}||_{\infty}}{||{\bfd}^{\mathrm{f},\alpha}_{n+p,k-1}||_{\infty}}  < \mathrm{tol}_\mathrm{f}.
    \end{equation*}
\end{algsubstates}
\State \textbf{If} \ref{state-error-substep-explicit-implicit} is met, \textbf{then} $\bfd^{\mathrm{f},\alpha}_{n+p} \gets \bfd^{\mathrm{f},\alpha}_{n+p,k}$, $\ddot{\bfd}^{\mathrm{f},\alpha}_{n+p} \gets \ddot{\bfd}^{\mathrm{f},\alpha}_{n+p,k}$, and $\ddot{\bfd}^{\mathrm{c}}_{n+p} \gets \ddot{\bfd}^{\mathrm{c}}_{n+p,k}$.
\State \textbf{Else} $k \gets k+1$, and \Goto{state-explicit-implicit-substep} for the next iteration.
\State The coarse-scale velocity is updated as in step \ref{state-eesm-velocity-substep} of Algorithm \ref{alg:EE-SSM}, and the fine-scale velocity update is:
    \begin{equation*}
        \dot{\bfd}^{\mathrm{f}, \alpha}_{n+p} = \left( \bfd^{\mathrm{f}, \alpha}_{n+p}  - \bfd^{\mathrm{f}, \alpha}_{n} \right) \frac{2}{p \Delta t}  - \dot{\bfd}^{\mathrm{f}, \alpha}_{n}; \hspace{0.3cm} \alpha=1,2,\dots,n_{\mathrm{es}}.
    \end{equation*}
\algstore{alg:EI-SSM}
\end{algorithmic}
\end{algorithm}
\begin{algorithm} [htb!]
\small
\ContinuedFloat 
\caption{Implementation of EI-SSM (cont.)}
\begin{algorithmic}[]
\algrestore{alg:EI-SSM}
\State \textbf{Full-step}:
\State Set $k=1$; Compute the external force vectors at $t_{n+1}$ and set the initial guesses: $\bfd^{\mathrm{f}, \alpha}_{n+1, 0}=\bfd^{\mathrm{f}, \alpha}_{n+p}$, $\ddot{\bfd}^{\mathrm{f}, \alpha}_{n+1, 0}=\ddot{\bfd}^{\mathrm{f}, \alpha}_{n+p}$.
\State At an arbitrary increment, $k>0$: \label{state-explicit-implicit-fullstep}
\begin{algsubstates}
    \State Update the coarse-scale accelerations as given in step \ref{state-eesm-displacement-fullstep} of Algorithm \ref{alg:EE-SSM}.
    \State Solve Eq.\eqref{eq:implicit-full-step-newton-raphson} iteratively for fine-scale displacement and acceleration, and the updates for the $i^\mathrm{th}$ iteration are shown below:
    \begin{align*}
        & \big( {\bfd}^{\mathrm{f}, \alpha}_{n+1,k} \big)^{(i)} = \big( {\bfd}^{\mathrm{f}, \alpha}_{n+1,k} \big)^{(i-1)} + \big( {\tilde{d} \bfd}^{\mathrm{f}, \alpha}_{n+1,k}\big)^{(i)}; \\ 
        & \big( \ddot{\bfd}^{\mathrm{f}, \alpha}_{n+1,k} \big)^{(i)} = c_3 \left( c_3 \big( \bfd^{\mathrm{f}, \alpha}_{n+1,k} \big)^{(i)} + c_2 \bfd^{\mathrm{f}, \alpha}_{n+p}  + c_1 \bfd^{\mathrm{f}, \alpha}_{n}\right) + c_2 \dot{\bfd}^{\mathrm{f}, \alpha}_{n+p}  + c_1 \dot{\bfd}^{\mathrm{f}, \alpha}_{n}.
    \end{align*}
    The Newton iterations are performed until the norm of the residual vector (the right-hand side of the Eq.~\eqref{eq:implicit-full-step-newton-raphson}) is below a set tolerance value.
    \State Check the following error quantities in addition to those mentioned in step \ref{state-error-cdm} of Algorithm \ref{alg:EE-CDM}: \label{state-error-fullstep-explicit-implicit}
    \begin{equation*}
        \frac{||{\bfd}^{\mathrm{f},\alpha}_{n+1,k} - {\bfd}^{\mathrm{f},\alpha}_{n+1,k-1}||_{\infty}}{||{\bfd}^{\mathrm{f},\alpha}_{n+1,k-1}||_{\infty}}  < \mathrm{tol}_{\mathrm{f}}; \hspace{0.3cm} \alpha=1,2,\dots,n_{\mathrm{es}}.
    \end{equation*}
\end{algsubstates}
\State \textbf{If} \ref{state-error-fullstep-explicit-implicit} is met, \textbf{then} $\bfd^{\mathrm{f},\alpha}_{n+1} \gets \bfd^{\mathrm{f},\alpha}_{n+1,k}$, $\ddot{\bfd}^{\mathrm{f},\alpha}_{n+1} \gets \ddot{\bfd}^{\mathrm{f},\alpha}_{n+1,k}$, and $\ddot{\bfd}^{\mathrm{c}}_{n+1} \gets \ddot{\bfd}^{\mathrm{c}}_{n+1,k}$.
\State \textbf{Else} $k \gets k+1$, and \textbf{go to} \ref{state-explicit-implicit-fullstep} for the next iteration.
\State The coarse-scale velocity is updated as in step \ref{state-eesm-velocity-fullstep} of Algorithm \ref{alg:EE-SSM} and the fine-scale velocity update is:
\begin{equation}
     \dot{\bfd}^{\mathrm{f}, \alpha}_{n+1} = c_3 \bfd^{\mathrm{f}, \alpha}_{n+1} + c_2 \bfd^{\mathrm{f}, \alpha}_{n+p}  + c_1 \bfd^{\mathrm{f}, \alpha}_{n}. \nonumber
\end{equation}
\State \Return $\big(\bfd^\mathrm{c}_{n+1}, \dot{\bfd}^\mathrm{c}_{n+1}, \ddot{\bfd}^\mathrm{c}_{n+1} \big)$ and $\big( {\bfd^{\mathrm{f}, \alpha}_{n+1}}, {\dot{\bfd}^{\mathrm{f}, \alpha}_{n+1}}, {\ddot{\bfd}^{\mathrm{f},\alpha}_{n+1}} \big)$.
\end{algorithmic}
\end{algorithm}

\subsection{Overall algorithm}
The overall algorithm for the semi-discrete multiscale equations is described in Algorithm \ref{algo:overall_algo_multiscale}. Given the initial displacement and velocity conditions at coarse- and fine-scales, the initial accelerations at both scales are obtained iteratively until convergence is achieved. For an arbitrary time step, the stable time increment is obtained based on the time integration method, and appropriate updates to the coarse- and fine-scale fields are performed.

\begin{algorithm} [htb!]
\small
\caption{{Algorithm for multiscale problem}} 
\label{algo:overall_algo_multiscale}
\begin{algorithmic}[1]
\State \textbf{Input}: Given an initial displacement and velocity condition for the coarse-scale problem $\big(\bfd^\mathrm{c}_0, \dot{\bfd}^\mathrm{c}_0 \big)$, along with loading and boundary conditions. The initial fine-scale displacement and velocity are $\big( {\bfd^{\mathrm{f}, \alpha}_0} = \mathbf{0}, {\dot{\bfd}^{\mathrm{f}, \alpha}_0} = \mathbf{0} \big)$. 
\State \textbf{Initial acceleration}: The initial acceleration for coarse-scale $\big( \ddot{\bfd}^\mathrm{c}_0 \big)$ and fine-scale  $\big( {\ddot{\bfd}^{\mathrm{f}, \alpha}_0} \big)$ problems are obtained from iteratively solving Eq.~\eqref{eq:ode_coarse_scale} and Eq.~\eqref{eq:ode_fine_scale}, until convergence is achieved. Set $n=0$.
\State Update for ${(n+1)}^{\mathrm{th}}$ time step:
\begin{algsubstates}
    \State  Known fields at $n^{\mathrm{th}}$ time step: coarse-scale $\left(\bfd^\mathrm{c}_n, \dot{\bfd}^\mathrm{c}_n, \ddot{\bfd}^\mathrm{c}_n \right)$ and fine-scale $\left( {\bfd^{\mathrm{f}, \alpha}_n}, {\dot{\bfd}^{\mathrm{f}, \alpha}_n}, {\ddot{\bfd}^{\mathrm{f},\alpha}_n} \right)$.
    \State Get a stable time increment $(\Delta t)$ for the multiscale problem as discussed in Sec.~\ref{sec:stable_time_increment_multiscale_problem}, which depends on the time integration method being employed.
    \State The coarse-scale $\left(\bfd^\mathrm{c}_{n+1}, \dot{\bfd}^\mathrm{c}_{n+1}, \ddot{\bfd}^\mathrm{c}_{n+1} \right)$ and fine-scale $\left( {\bfd^{\mathrm{f}, \alpha}_{n+1}}, {\dot{\bfd}^{\mathrm{f}, \alpha}_{n+1}}, {\ddot{\bfd}^{\mathrm{f},\alpha}_{n+1}} \right)$ using Algorithm \ref{alg:EE-CDM} for \textit{EE-CDM}, Algorithm \ref{alg:EE-SSM} for \textit{EE-SSM}, and Algorithm \ref{alg:EI-SSM} for \textit{EI-SSM}.
\end{algsubstates}
\State Set $n \gets n+1$, and repeat Step $3$ until desired.
\end{algorithmic}
\end{algorithm}


\subsection{Element matrices and vectors} \label{sec:element_matrices_and_vectors}
The evaluation of element-level matrices and vectors given in Eqs.~\eqref{eq:element-level-matrices-vectors-coarse}, \eqref{eq:element-level-matrices-vectors-fine} requires numerical integration of the appropriate entities that involve coarse-scale basis functions only, fine-scale basis functions only, and some involving both coarse and fine-scale basis functions. For brevity, the non-standard integration procedure for the internal force vector ($\bff^{\mathrm{c},\alpha_E}_{\mathrm{int}}$) in the coarse-scale problem is discussed below, and other non-standard element-level entities are evaluated similarly.
The internal force vector in the coarse-scale problem is shown below:
\begin{equation} \label{eq:force_vector_coarse_scale_example}
    \bff^{\mathrm{c},\alpha_E}_{\mathrm{int}}  = \int_{\Omega_{\alpha_E}} \left( \bfB^{\mathrm{c}}_{\alpha_E} \right)^T  [\bfP] (\bfX) \: dV  =  \sum_{e \in I_{\alpha_E}} \int_{\Omega_{\alpha_e}} \left( \bfB^{\mathrm{c}}_{\alpha_E} \right)^T  [\bfP] \: dV,
\end{equation}
where $I_{\alpha_E}$ is index set of fine-scale elements resolving the coarse-scale element, $\alpha_E$. The evaluation of each of the element-level entities at the fine scale in the summation given in Eq.~\eqref{eq:force_vector_coarse_scale_example} requires the interpolated values of the coarse-scale basis functions and their derivatives at the integration points of the fine-scale parent domain. This is not readily available as the coarse-scale functions are defined on the coarse-scale parent domain. To obtain the interpolated values, a two-scale mapping procedure is employed as discussed in Ref.~\cite{hu2020spectral}. This procedure involves first finding the coordinate of the integration point in the physical domain, using the fine-scale element isoparametric mapping, and then applying the coarse-scale element inverse isoparametric mapping to find the coordinates of the same integration point in the coarse-scale parent domain. The coordinate of the integration point of the fine-scale parent domain $(\bs{\xi}^{\mathrm{f}, \alpha_e})$ in the coarse-scale parent domain $(\bs{\xi}^{\mathrm{c}, \alpha_E})$ is given by:
\begin{equation}
    \bs{\xi}^{\mathrm{c}, \alpha_E} = \mathcal{M}^{-1}_{\mathrm{c}} \left( \mathcal{M}_{\mathrm{f}} \left( \bs{\xi}^{\mathrm{f}, \alpha_e} \right) \right)
\end{equation}
where $\calM_\mathrm{f}$ and $\calM_{\mathrm{c}}$ denote the fine-scale and coarse-scale isoparameteric mappings. 

\subsection{Critical time increment for multiscale and direct numerical simulations} \label{sec:stable_time_increment_multiscale_problem}
For direct numerical simulations of a nonlinear system of governing equations for structural dynamics, linearized stability analysis is performed to determine the critical time increment associated with a time integration method. Firstly, the generalized amplification matrix form is obtained by employing the updates in the time integration scheme to the semi-discrete equations. Then, the generalized amplification eigenvalue problem is decoupled into modal equations by expanding its eigenvectors in terms of the system eigenvectors $\bfK \bs{\Phi} = \omega^2_0 \bfM \bs{\Phi}$ \cite{belytschko2014nonlinear}. The critical time increment is obtained by restricting the moduli of the complex roots of the characteristic equation for the highest frequency mode in the uncoupled equations to be less than or equal to $1$. Noh and Bathe \cite{noh2013explicit} employed the Routh-Hurwitz stability criteria on the characteristic polynomial of the amplification matrix in decoupled modal equations for the explicit sub-step method to obtain a critical time increment as shown below: 
\begin{equation}
    \Delta t_{\mathrm{crit}} = \mathrm{CFL} \max_{I} \frac{2}{(\omega_0)_I},
\end{equation}
where $I$ is the index for eigenvalues $(\omega_0)_I$, and the maximum value of Courant–Friedrichs–Lewy $(\mathrm{CFL})$ allowed is $1/p$, where $p$ is the sub-step ratio. It is important to note that the explicit sub-step method \cite{noh2013explicit} has a higher stability limit than the explicit central difference method, for which the maximum value of $\mathrm{CFL}$ allowed is $1.0$.

In summary, for direct numerical simulations, the critical time increment is obtained in terms of the maximum eigenvalue of the system $\bfK \bs{\Phi} = \omega^2_0 \bfM \bs{\Phi}$. The maximum eigenvalue is estimated based on the element level eigenvalue problem, as the maximum absolute eigenvalue of the unconstrained system is upper-bounded by the maximum absolute element level eigenvalue \cite{belytschko1985stability}. Moreover, by the Rayleigh nesting theorem, the maximum eigenvalue of the assembled system with essential boundary conditions enforced is bounded by the maximum eigenvalue of the unconstrained element level eigenvalue problem \cite{belytschko2014nonlinear}. The estimate of the maximum eigenvalue for the element level eigenvalue problem in DNS for the examples considered here is discussed in Appendix \ref{sec:dns_eigenvalue}.

For the multiscale problem, similar ideas are employed to determine the critical time increments for both coarse- and fine-scale equations. As the operator-split procedure is used to solve coarse and fine-scale problems iteratively, the stable time increment for both problems can be deduced based on the linearized stability analysis of individual problems. 

The linearized weak forms of the coarse- and fine-scale PDEs are required for the stability analysis. The discrete form of the linearized weak form for the fine-scale problem is shown in Eq.~\eqref{eq:discretized_linear_weak_form}. Following a similar procedure, the discretized form of the linearized weak form for the coarse-scale PDE is obtained as follows:
\begin{equation}
  \bfM^{\mathrm{c}} \tilde{d}\ddot{\bfd}^\mathrm{c} +  \bfK^{\mathrm{c}} \tilde{d}\bfd^\mathrm{c}  =  -  \bfM^{\mathrm{c}} \ddot{\bfd}^\mathrm{c} - \sum_{\alpha=1}^{n_{\mathrm{es}}} \bfM^{\mathrm{c} \mathrm{f}_{\alpha}} \ddot{\bfd}^{\mathrm{f}_\alpha} -  \bff^{\mathrm{c}}_{\mathrm{int}} \left( \bfd^\mathrm{c}, \bfd^{\mathrm{f}} \right) + \bff^\mathrm{c}_{\mathrm{ext}},
\end{equation}
where $\bfK^{\mathrm{c}}$ is the coarse-scale stiffness matrix, and $(\tilde{d}\bfd^\mathrm{c}, \tilde{d}\ddot{\bfd}^\mathrm{c})$ denote the discrete perturbations in coarse-scale fields. $\bfK^{\mathrm{c}}$ can be obtained using the element-level contributions as shown below:
\begin{equation}
\begin{aligned}
     \bfK^{\mathrm{c}} =  \sum_{\alpha, E} \left( \bfL^\mathrm{c}_{\alpha_E}\right)^T \bfK^{\mathrm{c}}_{\alpha_E} \bfL^\mathrm{c}_{\alpha_E} , \quad \text{where} \quad 
    \bfK^{\mathrm{c}}_{\alpha_E} = \sum_{e \in I_{\alpha_E}} \int_{\Omega_{\alpha_e}} \left( \bfB^\mathrm{c}_{\alpha_E} \right)^T  [\bfD] \bfB^\mathrm{c}_{\alpha_E}  \: dV.
\end{aligned}    
\end{equation}
 Following the procedure discussed for DNS, one can obtain the critical time increments for linearized coarse-scale and fine-scale systems given by:
\begin{equation}
\begin{aligned}
    & \text{coarse-scale system-} \quad \bfK^{\mathrm{c}} \bs{\Phi}^\mathrm{c} = \left(\omega^\mathrm{c}_0\right)^2 \bfM^{\mathrm{c}} \bs{\Phi}^\mathrm{c}, \\
    & \text{fine-scale system-} \quad \bfK^{\mathrm{f}_{\alpha}} \bs{\Phi}^{\mathrm{f}_{\alpha}} = \left(\omega^{\mathrm{f}_{\alpha}}_0\right)^2 \bfM^{\mathrm{f}_{\alpha}} \bs{\Phi}^{\mathrm{f}_{\alpha}}. 
\end{aligned}
\end{equation}
The critical time increments for the coarse-scale and fine-scale problems are as follows:
\begin{equation}
\begin{aligned}
     & \Delta t^\mathrm{c}_{\mathrm{crit}} = \mathrm{CFL} \max_{I} \frac{2}{ 
 (\omega^\mathrm{c}_0 )_I}, \\
    & \Delta t^{\mathrm{f}_{\alpha}}_{\mathrm{crit}} = \mathrm{CFL} \max_{\alpha} \left( \max_{I} \frac{2}{ (\omega^{\mathrm{f}_{\alpha}}_0 )_I} \right) .  \end{aligned}
\end{equation}
The maximum eigenvalues for the coarse-scale and fine-scale problems are estimated based on the corresponding element-level problems, similar to DNS. The estimate of the maximum eigenvalue for the element-level eigenvalue problem in coarse and fine-scale differential equations in VME simulations for the examples considered here is discussed in Appendix \ref{sec:vme_eigenvalue}. It is expected that the stable time increment of the coarse-scale problem will be larger than that of the fine-scale problem, as $ \omega^{\mathrm{f}_{\alpha}}_0 >> \omega^\mathrm{c}_0$. Hence, the stable time increment for the multiscale problems based on the integration schemes can be deduced as follows:
\begin{itemize}
    \item For EE-CDM or EE-SSM algorithm, $\Delta t_{\mathrm{crit}} = \Delta t^{\mathrm{f}_{\alpha}}_{\mathrm{crit}}$. The $\mathrm{CFL}$ is chosen based on the integration scheme employed.

    \item For the EI-SSM algorithm, it is given by $\Delta t_{\mathrm{crit}} = \Delta t^\mathrm{c}_{\mathrm{crit}}$, as the implicit update for the fine-scale problem is unconditionally stable. 
\end{itemize}

%
\vspace{-0.3cm}
\section{Numerical Verification} \label{sec:numerical_verification}

This section presents numerical examples of wave propagation in a one-dimensional domain modeled using the compressible Neo-Hookean material model (see Appendix \ref{sec:neo-hookean-model}). While the numerical schemes developed in this manuscript apply to multidimensional problems, one-dimensional cases directly address the issues associated with multiscale time integration stability. We therefore restricted the numerical analysis to one-dimensional cases. First, the performance of time integration methods, including the explicit central difference method, explicit-explicit sub-step method, and explicit-implicit sub-step method, is evaluated for wave propagation in VME simulations for a homogeneous microstructure. Next, the effect of contrast in the elastic modulus of a heterogeneous microstructure on wave propagation is examined. Then, the effect of different initial displacement fields on wave propagation is examined in multiscale simulations, demonstrating that the VME approach accurately reproduces key wave propagation characteristics such as dispersion and attenuation, consistent with DNS results. Finally, the computational performance of explicit-explicit and explicit-implicit time integration schemes is assessed under varying contrasts in the elastic modulus of a heterogeneous microstructure.

The original equation of motion in the 1-D case is expressed in the following form:
\begin{equation} \label{eq:1-D_lmb}
    \frac{\partial P(X,t)}{\partial X} = \rho_0 (X) \frac{\partial^2 u(X,t)}{\partial t^2},
\end{equation}
where $P$ denotes the first Piola-Kirchhoff stress, and $u$ is the displacement. The non-dimensional form of Eq.~\eqref{eq:1-D_lmb} is obtained by introducing the following entities:
\begin{equation} \label{eq:nondimensional_entitites}
    \tilde{X} = \frac{X}{L}; \quad \tilde{t} = \frac{v \: t}{L}; \quad \tilde{P} = \frac{P}{E^A}; \quad \tilde{u} = \frac{u}{L},
\end{equation}
where $L$ is the length of the domain, $v = \sqrt{{E^A}/{\rho^A_0}}$ is the wave speed of material $A$, $E^A$ is the Young's modulus of material $A$, and $\rho^A_0$ is the mass density of material $A$ in the reference configuration. Substituting Eq.~\eqref{eq:nondimensional_entitites} into Eq.~\eqref{eq:1-D_lmb}, we obtain the non-dimensional form of the equation of motion:
\begin{equation}
    \frac{\partial \tilde{P}(\tilde{X},\tilde{t})}{\partial \tilde{X}} = \frac{\rho_0(\tilde{X})}{\rho^A_0} \frac{\partial^2 \tilde{u}(\tilde{X},\tilde{t})}{\partial \tilde{t}^2}.
\end{equation}
The corresponding multiscale system is derived as described in Section~\ref{sec:VME_Formulation}. In what follows, the non-dimensional form of the governing equation is solved for all numerical examples demonstrated below. The $\tilde{\left( \cdot \right)}$ symbol is omitted from the non-dimensional entities for simplicity of the presentation. 

The initial conditions considered for all the examples demonstrated below are as follows ($X\in[-1/2,1/2]$):
\begin{equation} \label{eq:initial_displacement_1d}
\begin{aligned}
    &  u(X,0) = a \left( 1 - \tanh^2 \left(\frac{X}{c} \right) \right), \\
    & \dot{u}(X,0) = 0,
\end{aligned}
\end{equation}
and, the boundary conditions are $u(-1/2, t) = 0$ and $u(1/2, t) = 0$, unless stated otherwise.

For all multiscale simulations (referred to as VME below) reported in this work, unless otherwise mentioned, the 1-D domain is discretized using $n_\mathrm{es} = 100$ unit cells, $n_{\mathrm{ecp}} = 1 $ coarse element per unit cell,  and $n_\mathrm{ef} = 8$ fine-scale quadratic elements for each unit cell. The results of the multiscale simulations are compared with the direct numerical simulations (DNS), which are obtained using the finite element method, where the material microstructure is resolved throughout the problem domain. To be consistent, the domain is discretized into $800$ quadratic elements for DNS. Further refinement of the domain does not result in significant accuracy improvements in the cases discussed below. The sub-step ratio for the EE-SSM or EI-SSM of time integration in VME simulations and explicit sub-step integration in DNS is taken to be $p=0.54$, as suggested in Ref.~\cite{noh2013explicit}. The tolerance value for convergence between coarse and fine-scale problems in VME simulations is 1E-3 for all time-integration methods. The tolerance value for the Newton-Raphson iterations in the fine-scale problem for implicit updates at both sub-step and full step is 1E-10.

\subsection{Homogeneous domain} \label{sec:homogeneous_domain}
In this section, we assess the performance of different time integration schemes for wave propagation for both VME and DNS simulations. The initial displacement profile is obtained using Eq.~\eqref{eq:initial_displacement_1d} with $a=0.04$ and $c=0.05$.  

\begin{figure}[htb!]
    \centering
    \subfloat[][]{
    \includegraphics[scale=0.45]{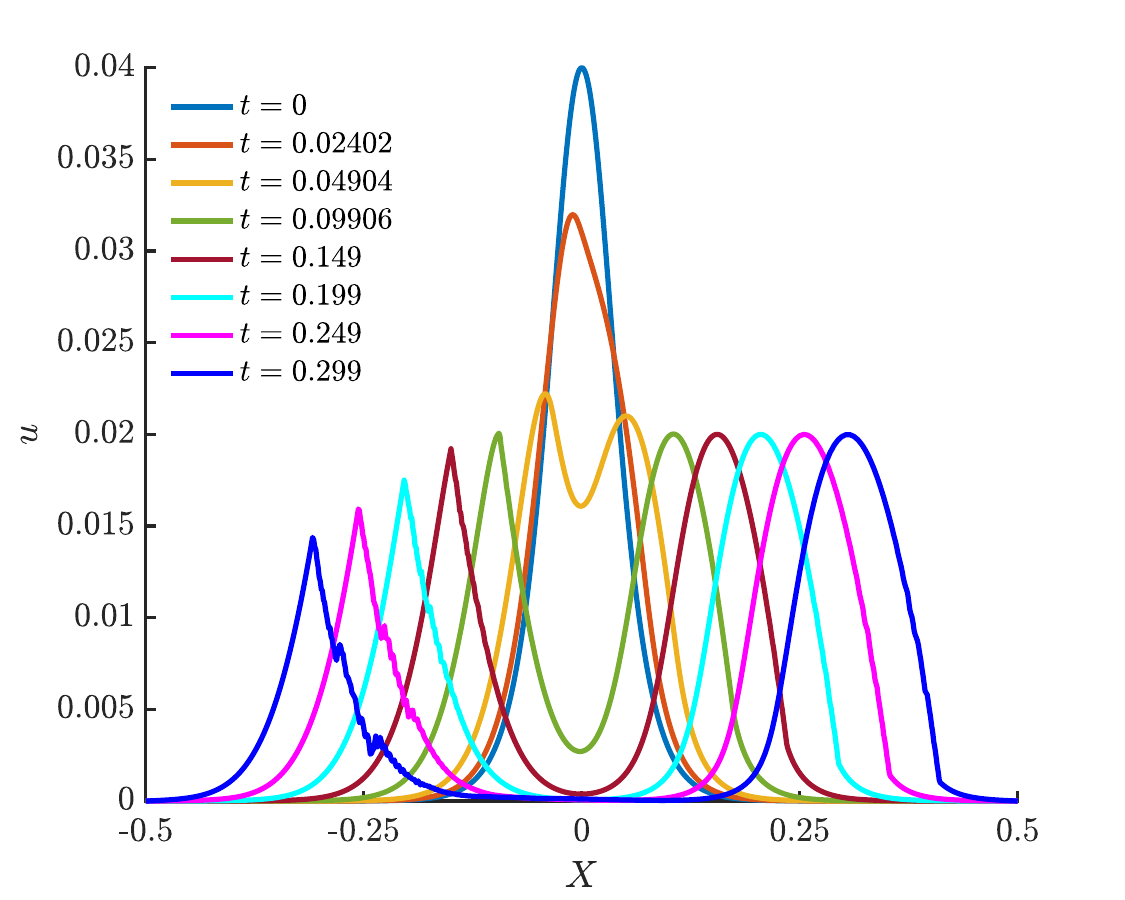}}
    \subfloat[][]{
    \includegraphics[scale=0.45]{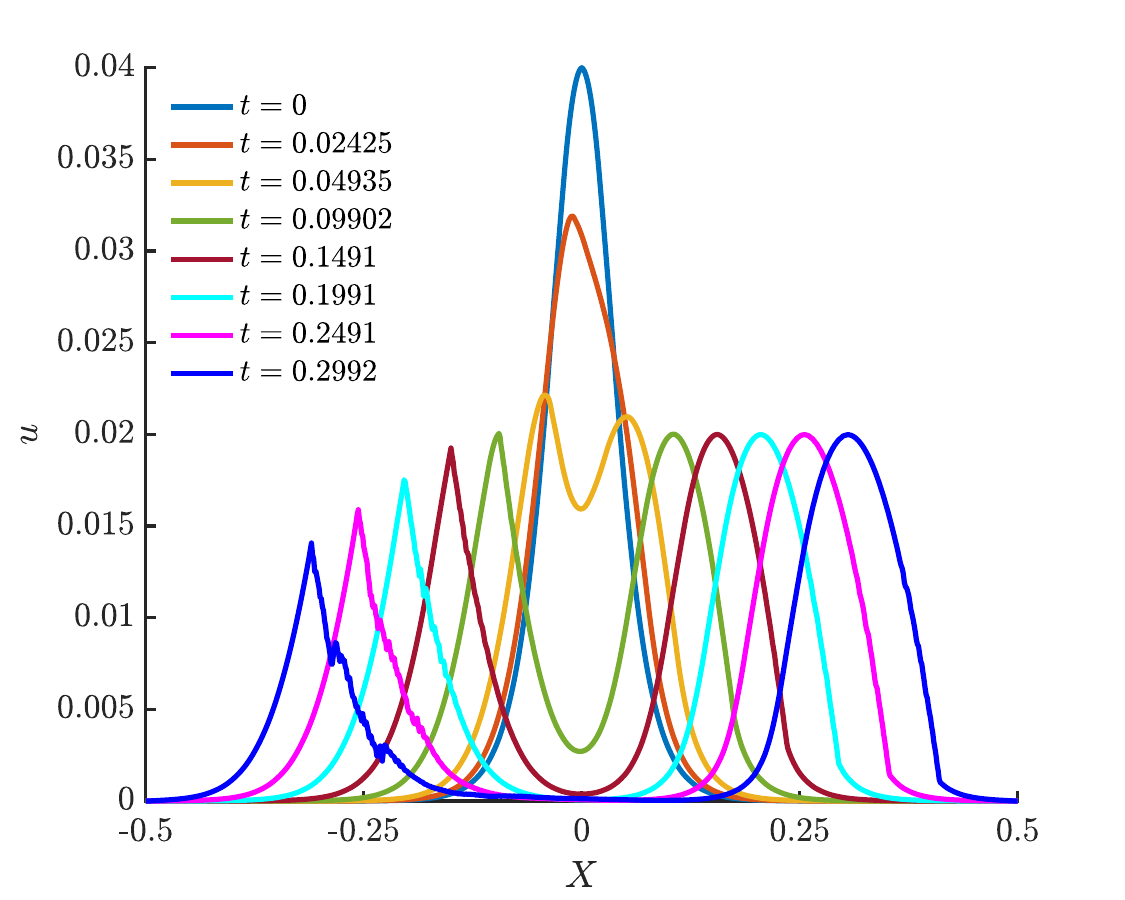}}
    \caption{Evolution of displacement with time predicted using the explicit central difference methods for the homogeneous material case with (a)~DNS and (b)~VME methods.}
    \label{fig:vme_dns_comparison_explicit_cdm}
\end{figure}

\begin{figure}[htb!]
    \centering
    \includegraphics[scale = 0.45]{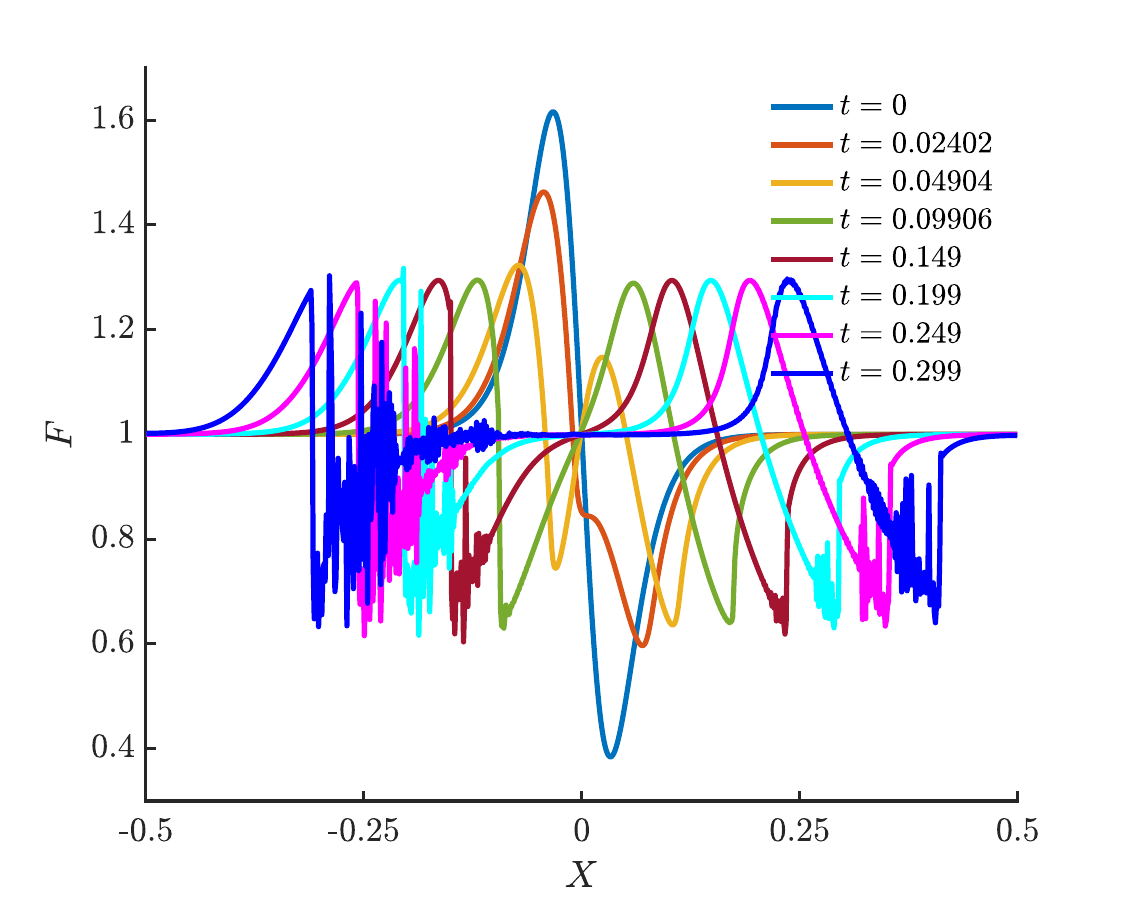}
    \caption{Averaged stretch over the element for the homogeneous material case in DNS.}
    \label{fig:dns_homogeneous_deformation_gradient}
\end{figure}

We first consider the homogeneous microstructure within all coarse-scale elements, by which $\rho_0/\rho^A_0 = 1$ and $E/E_A = 1$. Figure~\ref{fig:vme_dns_comparison_explicit_cdm} shows the evolution of the total displacement profile for a homogeneous microstructure using the explicit central difference method for both DNS and VME (using the EE-CDM approach) simulations. The results of the VME simulations are reported in the total form where the fine- and coarse-scale parts of the solution are evaluated separately and summed. Both the DNS and VME simulation results show that the initial displacement profile induces two waves traveling in opposite directions, consistent with the D'Alembert solution for linear wave propagation. The evolution of (element-averaged) stretch along the $X$ direction given by $F = 1 + du/dX$, is shown in Fig.~\ref{fig:dns_homogeneous_deformation_gradient}. Material points with $F>1$ indicate stretching, while those with $F<1$ indicate compression.

Due to the geometric and material nonlinearities in the Neo-Hookean model, the local wave speed becomes amplitude-dependent, leading to wave steepening. Specifically, the wave speed is lower in the tensile region than in the compressive region as the tangent modulus in compression is larger than in tension, for the same magnitude of displacement gradient (see Eq.~\eqref{eq:wave_speed} in Appendix \ref{sec:dns_eigenvalue}), resulting in asymmetric propagation. As shown in Fig.~\ref{fig:vme_dns_comparison_explicit_cdm}, this asymmetry manifests as a narrower crest on the tensile (left) side and a broader crest on the compressive (right) side, along with a reduction in amplitude on the tensile side. The compressive part of the traveling wave on the left side travels faster than the tensile part, leading to a reduction of the peak of the displacement wave as a function of time. Spurious oscillations appear in the displacement and strain wave profiles due to numerical dispersion errors introduced by the central difference method, particularly affecting high-frequency components of the solution. The numerical dispersion effects are apparent after $t=0.249$ for both DNS and VME simulations.

\begin{figure}[htb!]
    \centering
    \subfloat[][]{
    \includegraphics[scale=0.45]{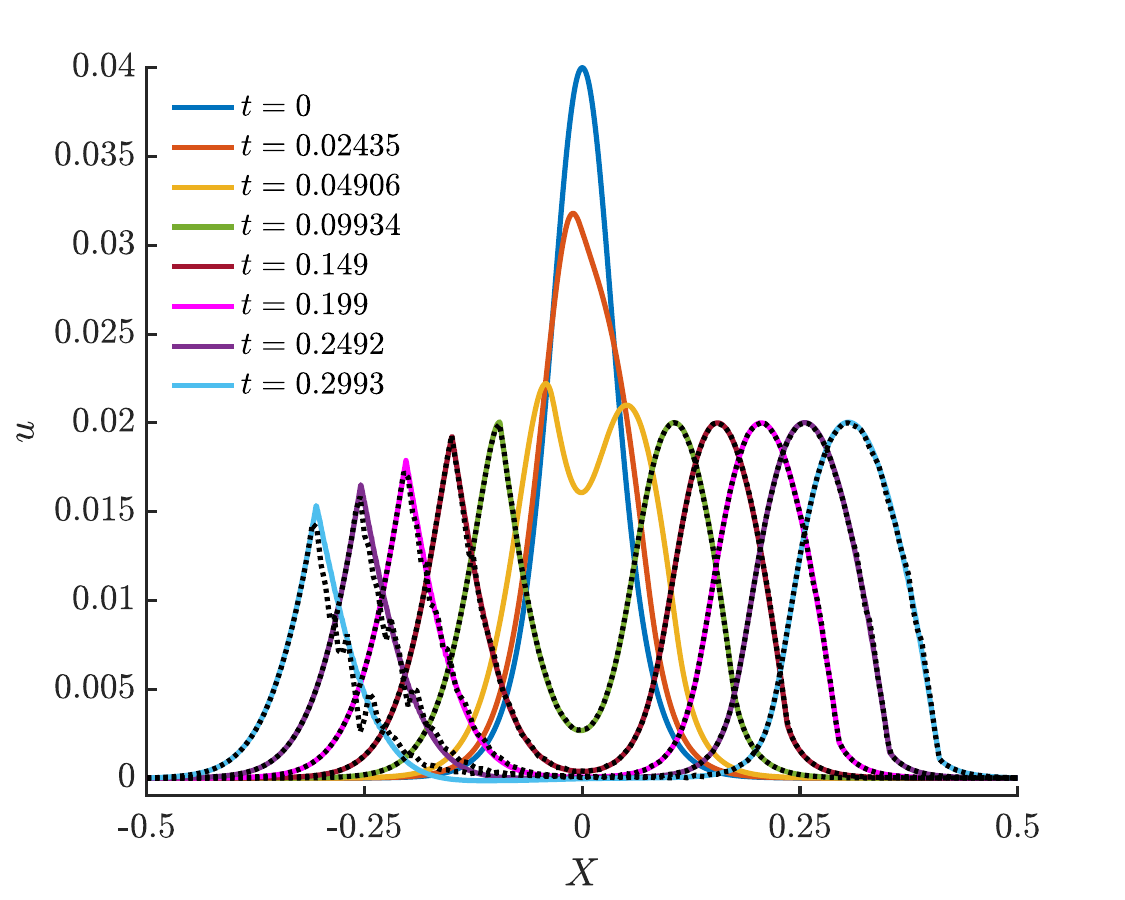}}
    \subfloat[][]{
    \includegraphics[scale=0.45]{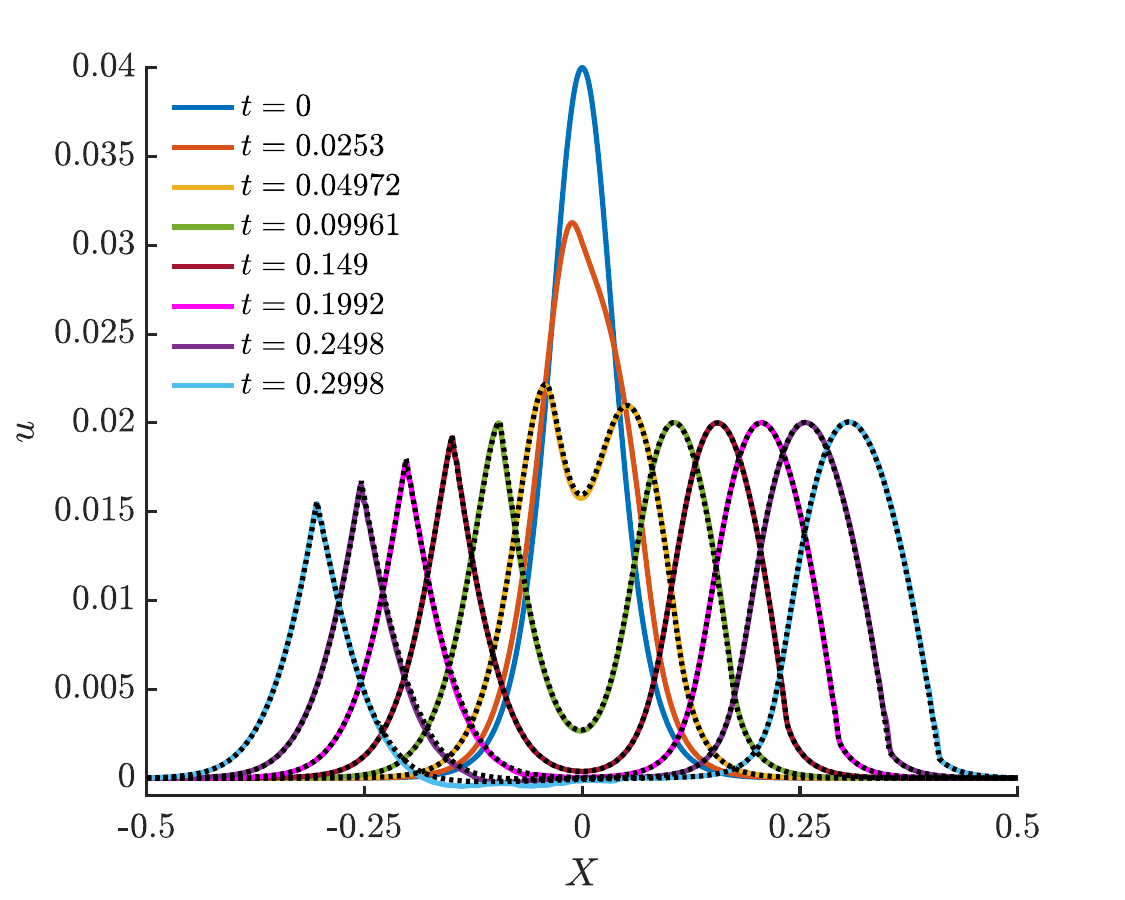}}
    \caption{Evolution of displacement profile with time for homogeneous microstructures using (a) EE-SSM, (b) EI-SSM in VME simulations. In both figures, the colored solid lines correspond to VME results, while the dotted lines in (a) correspond to the EE-SSM-based VME with coarse-grid approximation alone, and in (b) correspond to DNS, at the same non-dimensional times as the colored solid lines.}
    \label{fig:vme_dns_comparison_bathe_explicit_implicit}
\end{figure}

Figure \ref{fig:vme_dns_comparison_bathe_explicit_implicit} shows a comparison of the displacement waves predicted by the VME and DNS approaches using dissipative integration schemes. The DNS simulation employs the explicit sub-step integration method proposed in Ref.~\cite{noh2013explicit}, whereas for VME simulations, the EE-SSM and EI-SSM are utilized. $\mathrm{CFL} = 1$ is used for both DNS and EE-SSM-based VME simulations, while $\mathrm{CFL} = 0.5$ is taken for EI-SSM-based VME simulations. As evident in Fig.~\ref{fig:vme_dns_comparison_bathe_explicit_implicit}, spurious oscillations do not develop when the dissipative methods are employed. This is because the contribution of high-frequency modes to the overall solution is reduced in these methods by decreasing the spectral radius of the amplification matrix for shorter wavelengths \cite{bathe2005composite, noh2013explicit}. In Fig.~\ref{fig:vme_dns_comparison_bathe_explicit_implicit}, the relative error based on the $L^{\infty}$ norm of the total displacement field between the VME and DNS simulations is $0.0051$ for EE-SSM at $t=0.2993$, and $0.0259$ for EI-SSM at $t=0.2998$. Hence, for the same discretization, the EE-SSM and EI-SSM-based VME simulations remove the high-frequency oscillations that are present for the CDM-based VME simulations. We note that, unlike computational homogenization-based methods, the fine-scale solution in VME is not necessarily induced by heterogeneity in the microstructure. In the case of a homogeneous domain, the fine-scale solution captures the discretization errors induced by the coarse-scale grid and effectively improves the accuracy of the solution. Figure~\ref{fig:vme_dns_comparison_bathe_explicit_implicit}a includes the results obtained using the coarse grid approximation alone, which deviates from the multiscale and the DNS solutions, especially at later times, where the relative error of the total displacement compared to DNS is $0.1504$ at $t= 0.2993$. This is due to the accumulation of large numerical dispersion errors with coarse-grid approximation alone in the VME simulation.

\subsection{Heterogeneous domain}
In this section, we study the effect of contrast in the elastic modulus in a heterogeneous microstructure on wave propagation. Let the domain consist of a repeated two-phase microstructure such that:
\begin{equation} \label{eq:elastic_modulus_hetero}
E = \left\{
\begin{aligned}
   &&  E^A, \quad \text{if} \quad X \in \Omega_A; \\
   &&  C \: E^A, \quad \text{if} \quad X \in \Omega_B; 
\end{aligned}   
\right.
\end{equation}
where $C$ is the modulus contrast ratio, $\Omega_A = [-kl -L/2,(k+\beta) l - L/2)$ and $\Omega_B = [(k+\beta) l -L/2, (k+1)l -L/2)$, with $l$ the size of the microstructure, $\beta$ the fraction of material $A$, and $k \in \{ 0,1,\ldots,(L/l)-1\}$. The impedance mismatch is taken to be generated by the modulus contrast alone, and the mass density in the reference configuration is taken as constant, i.e.~$\rho_0 = \rho^A_0$. 


\begin{figure}[htb!]
    \centering
    \subfloat[][]{
    \includegraphics[scale=0.45]{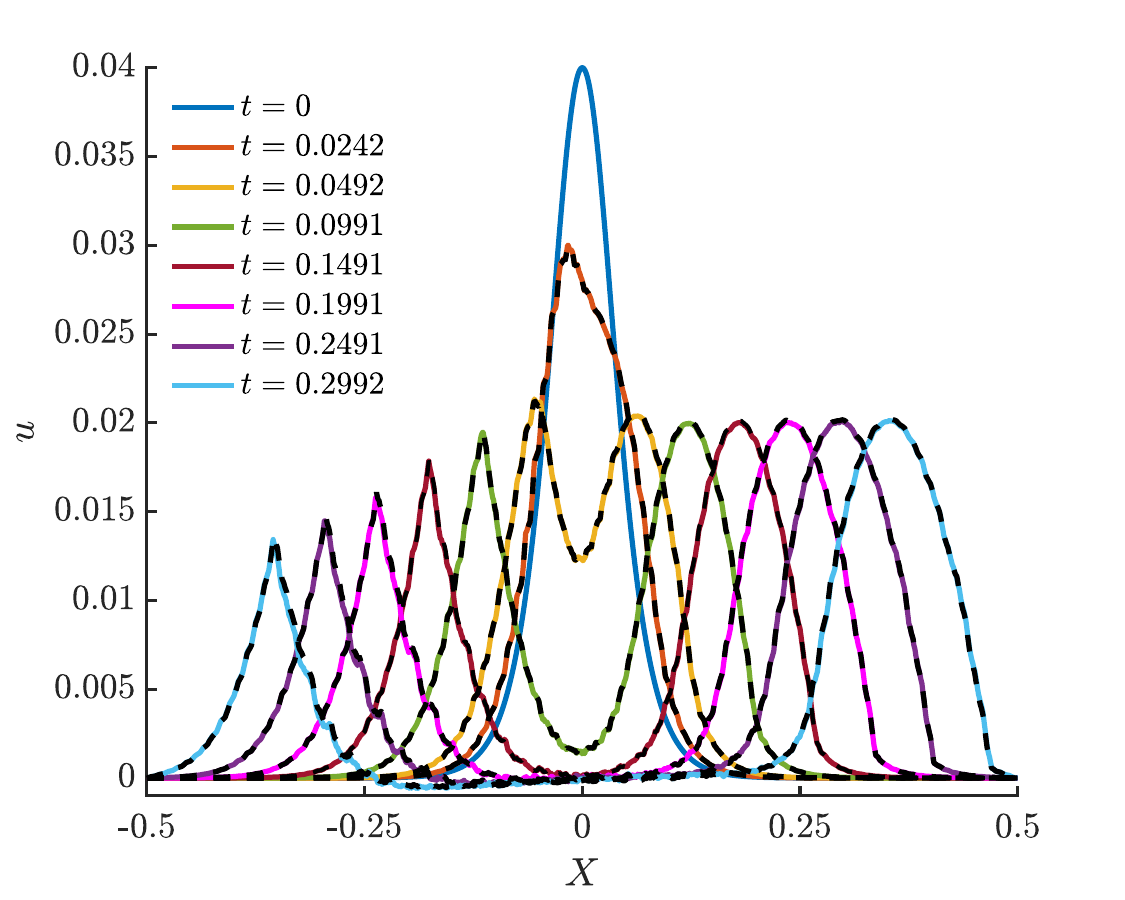}}
    \subfloat[][]{
    \includegraphics[scale=0.45]{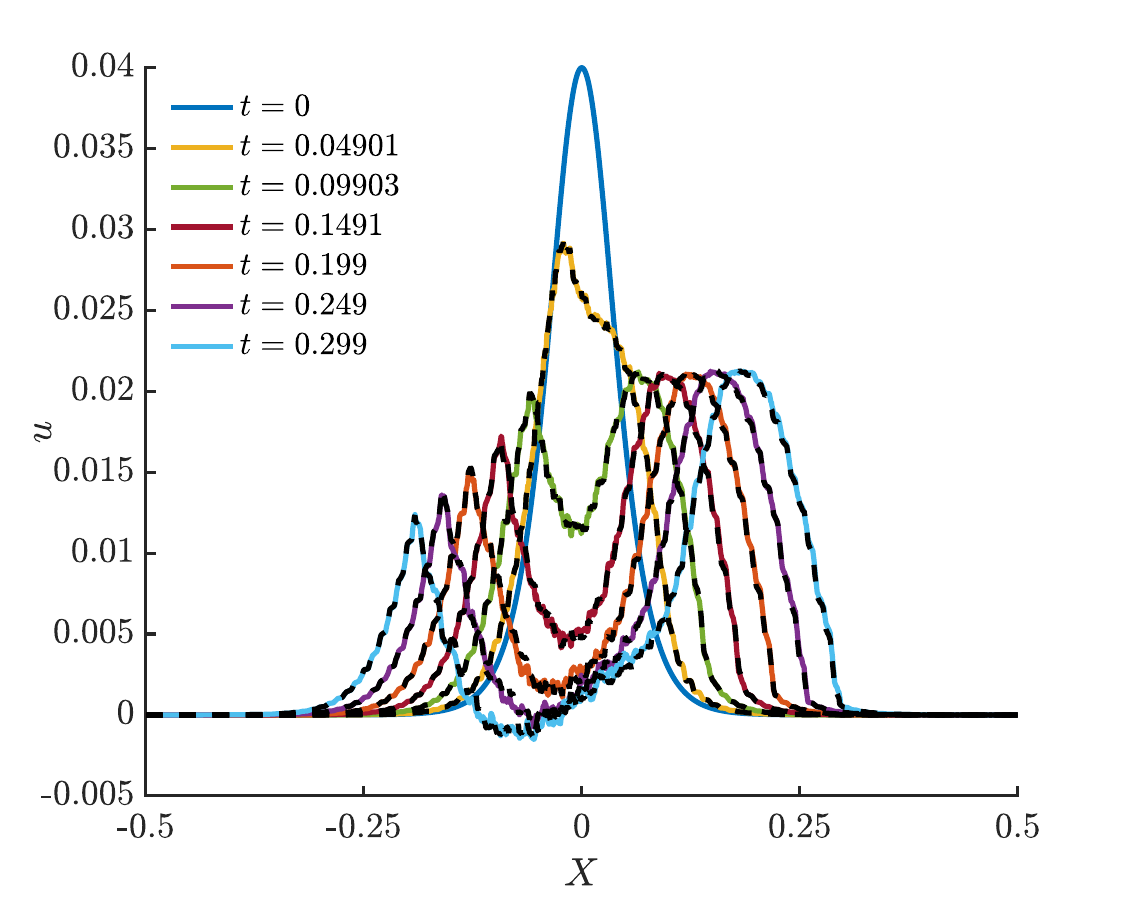}}
    \caption{Evolution of displacement profile with time using EE-SSM for VME and explicit sub-step method for DNS simulations of heterogeneous microstructures (a) $C=2$, (b) $C=0.2$. In both figures, the colored solid lines correspond to VME results, and the black dashed lines correspond to the DNS results.}
    \label{fig:vme_dns_comparison_bathe_explicit_heterogeneous}
\end{figure}

Figures~\ref{fig:vme_dns_comparison_bathe_explicit_heterogeneous}a-b show the propagation of the displacement wave induced by the initial displacement profile obtained using Eq.~\ref{eq:initial_displacement_1d} with $a=0.04$ and $c=0.05$ for $C=2$ (i.e., phase contrast of 2) and $C=0.2$ (phase contrast of 5), respectively. The microstructure parameters are set to $l=0.01$, $\beta=0.5$, and each unit cell is associated with one coarse-scale element i.e., $n_\mathrm{ecp} =1$. 
The DNS simulations were performed using the explicit sub-step integration method, whereas the VME simulations employ EE-SSM for time integration. A $\mathrm{CFL}$ number of $1$ is used for $C=2$, while $\mathrm{CFL} = 0.2$ is used for $C=0.2$. The $\mathrm{CFL}$ was reduced for the $C=0.2$ case, as it was observed that using larger values of $\mathrm{CFL}$ led to numerical instability. This can happen in nonlinear problems, as the stable time increment is obtained from linearized stability analysis, and the actual response may deviate from the linear approximation. As observed in Fig.~\ref{fig:vme_dns_comparison_bathe_explicit_heterogeneous}, the displacement waves travel faster or slower depending on the increase/decrease in the elastic modulus compared to the homogeneous microstructure. Due to contrast in the elastic modulus, especially for $C=0.2$, wave reflections are more prominent, leading to oscillations in the wave profile with evolution in time. However, no significant wave dispersion arises from the microstructural heterogeneity, as the initial wavelength is substantially larger than the microstructural length scale. As a result, the propagating wave interaction with the microstructure is limited. The VME simulation captures the resulting oscillations with reasonable accuracy compared to the DNS results, with the relative error in the total displacement compared to DNS being $0.02496$ for $C=2$, and $0.1218$ for $C=0.2$, both at $t=0.299$.

\begin{figure}[htb!]
    \centering
    \includegraphics[scale = 0.45]{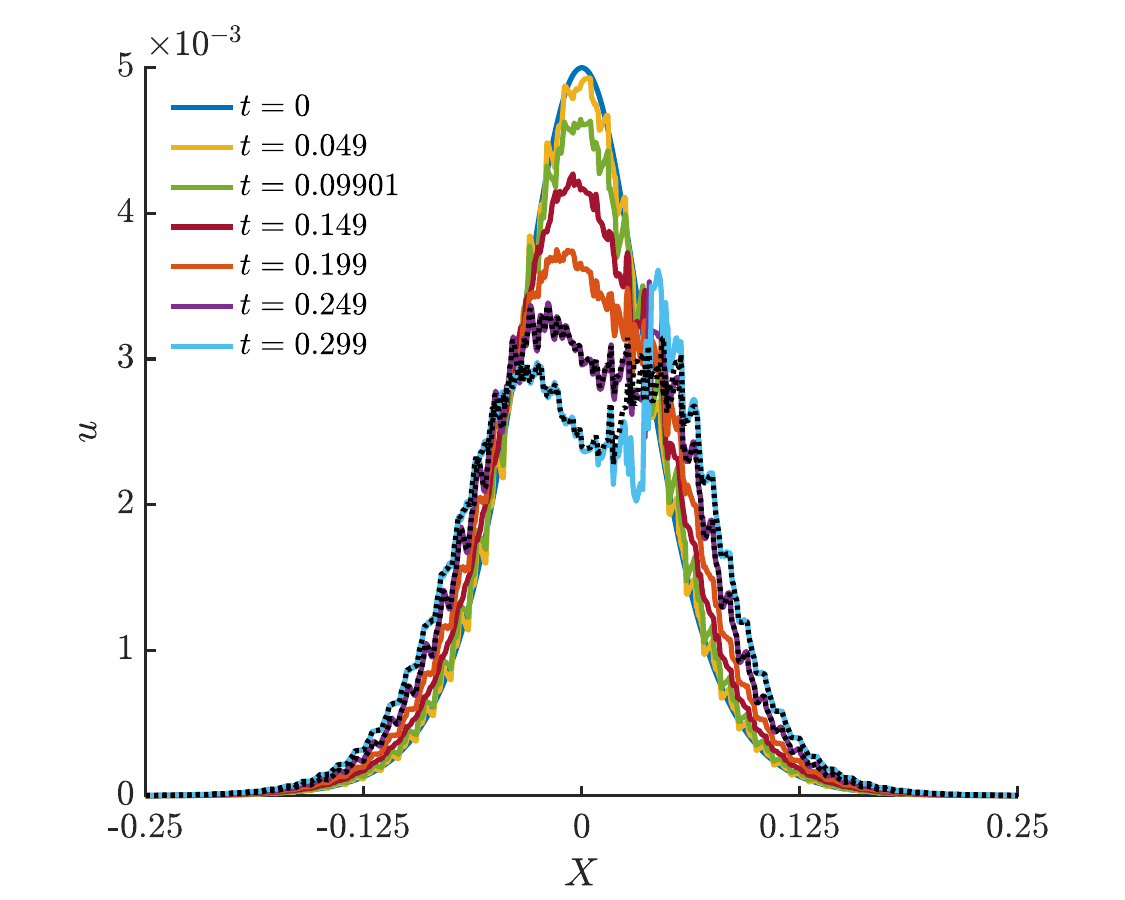}
    \caption{Evolution of displacement profile with time using EE-SSM for VME simulations for a heterogeneous microstructure with $C=0.01$ and $\mathrm{CFL} = 0.1$. The dashed black lines show DNS results at the same non-dimensional times using the explicit sub-step method.}
    \label{fig:VME_DNS_heterogeneous_C_0_01_CFL_0_1}
\end{figure}

Figure~\ref{fig:VME_DNS_heterogeneous_C_0_01_CFL_0_1} shows the evolution of the displacement field for a heterogeneous microstructure with $C=0.01$ (phase contrast of 100) using EE-SSM for VME simulations and explicit sub-step method for DNS. The initial displacement profile is chosen with $a=0.005$ in Eq.~\eqref{eq:initial_displacement_1d}, and this is because if larger amplitudes are taken, then it results in unphysical compressive strains with stretch, $F < 0.01$. The $\mathrm{CFL} = 0.1$ is taken for both VME and DNS results. The relative error in the displacement field obtained from VME compared to DNS is $0.27138$ at $t=0.299$, and this is resulting from a relatively larger tolerance chosen for convergence in iterations between coarse- and fine-scale problems.

\begin{figure}[htb!]
    \centering
    \subfloat[][]{
    \includegraphics[scale = 0.45]{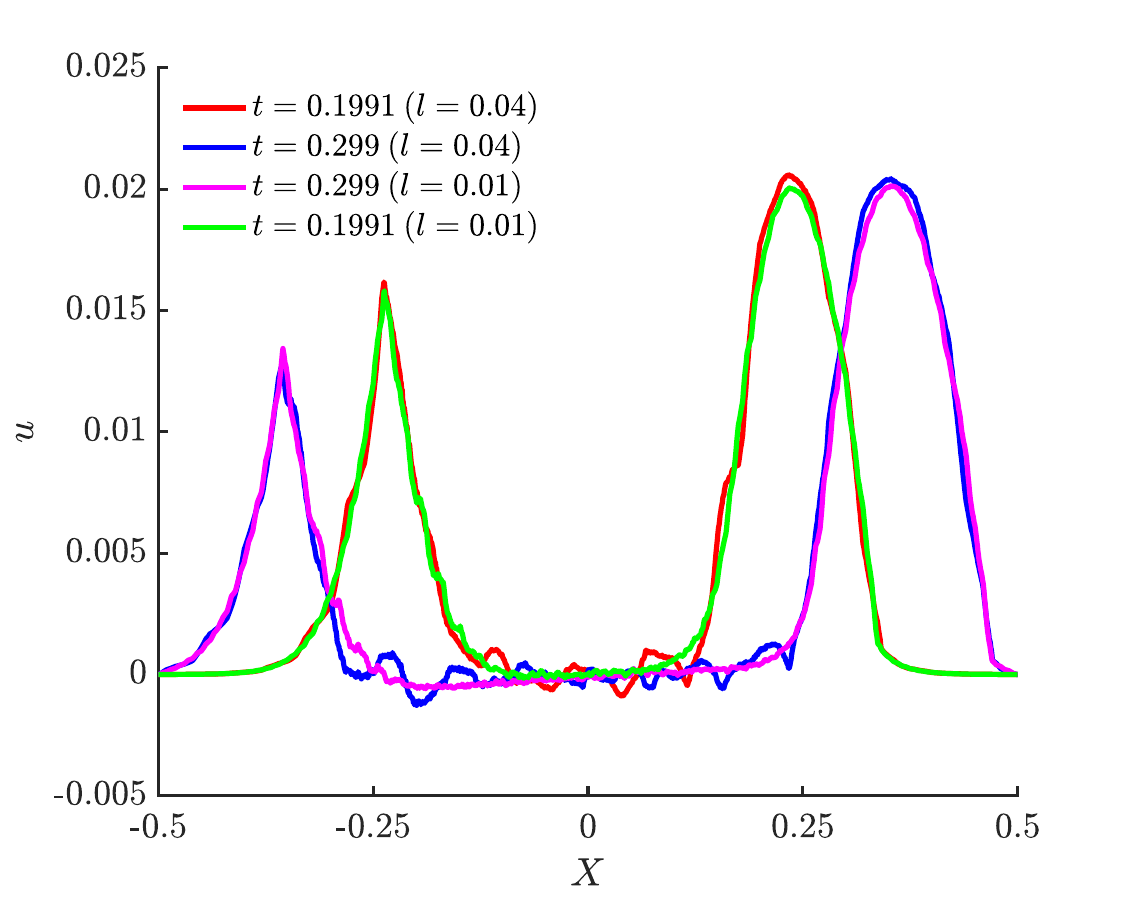}}
    \subfloat[][]{
    \includegraphics[scale = 0.45]{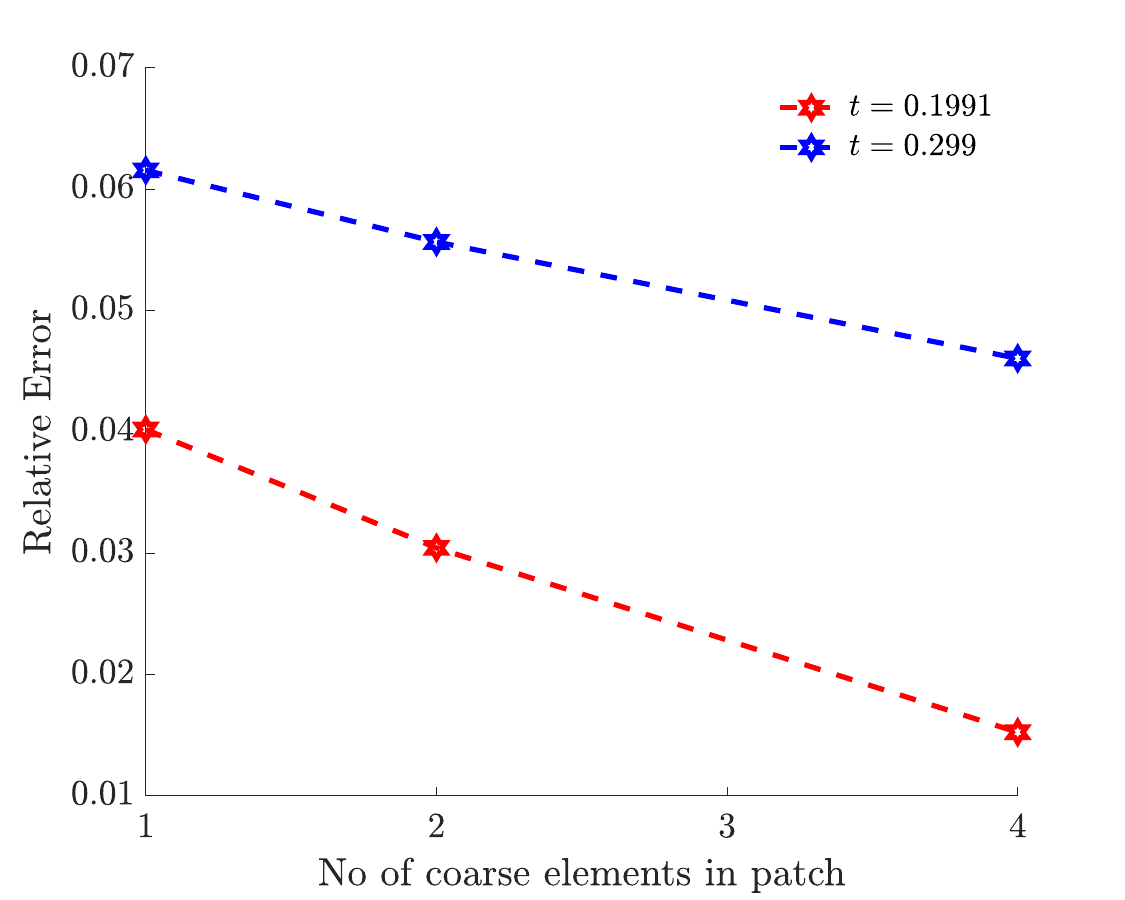}}
    \caption{Wave dispersion using EE-SSM for VME simulations of heterogeneous microstructure with $C=2$, $n_\mathrm{es} = 25$, and $n_{\mathrm{ef}} = 32$. (a) $n_{\mathrm{ecp}} = 4$ for $l=0.04$, and $n_{\mathrm{ecp}} = 1$ for $l =0.01$, (b) comparison of relative error with respect to DNS as a function of $n_{\mathrm{ecp}}$ for $l=0.04$.}
    \label{fig:vme_dns_comparison_dispersion_heterogeneous}
\end{figure}

To investigate wave dispersion effects, the size of the microstructure is increased, thereby promoting the interaction between the propagating wave and the material heterogeneity. We consider $l=0.04$, i.e., $n_\mathrm{es} = 25$ microstructures in the domain, and use a fine-scale grid $n_{\mathrm{ef}} = 32$ per unit cell, and three coarse-scale grids are considered: (a) $n_{\mathrm{ecp}} = 1$, (b) $n_{\mathrm{ecp}} = 2$, and (c) $n_{\mathrm{ecp}} = 4$. The error in the wave propagation is evaluated as a function of the number of elements in a patch of coarse-scale elements that discretize a unit cell. Figure~\ref{fig:vme_dns_comparison_dispersion_heterogeneous}a shows the evolution of the displacement profiles for smaller ($l=0.01$, $n_{\mathrm{ecp}} = 1$) and larger ($l=0.04$, $n_{\mathrm{ecp}} = 4$) microstructures. The VME simulations employ the EE-SSM time integration scheme with a CFL number of $0.5$. When the microstructural size increases, the propagating displacement wave interacts more strongly with the microstructure, and dispersive effects become significant. For example, the left-traveling peak is smaller for $l=0.04$ compared to $l=0.01$ -- a consequence of dispersion. Similarly, on the right side, the broadening of the wave profile further confirms the presence of dispersive behavior. Figure~\ref{fig:vme_dns_comparison_dispersion_heterogeneous}b shows the relative error for $l=0.04$ compared to DNS for an increasing number of elements in a patch of coarse elements, i.e., $n_{\mathrm{ecp}}$. The DNS were obtained with $n_{\mathrm{el}} = 800$ elements with a $\mathrm{CFL} = 0.5$. As expected, the relative errors decrease with an increase in $n_{\mathrm{ecp}}$ and also increase with time. This will become even more critical as the microstructural size increases, particularly in higher-dimensional problems.


\subsection{Effect of initial displacement conditions}
In this section, we examine how the initial displacement profile influences wave propagation. In particular, we focus on the role of wavelength, controlled by the parameter $c$ in Eq.~\eqref{eq:initial_displacement_1d}, for a heterogeneous microstructure. The effect of amplitude on the solution has already been demonstrated through its deviation from the D’Alembert solution as discussed in Section~\ref{sec:homogeneous_domain}. For the heterogeneous case, the microstructure is defined with a phase contrast of $2$ (by setting $C = 0.5$), size $l= 0.01$, and the coarse-scale mesh with $n_\mathrm{ecp} = 1$. The initial displacement field is specified with $a=0.01$, $c=0.01$ in Eq.~\eqref{eq:initial_displacement_1d}. To distinguish the roles of nonlinearity and heterogeneity, we also compare the results with those of the homogeneous problem. 

\begin{figure}[htb!]
    \centering
    \subfloat[][]{
    \includegraphics[scale=0.45]{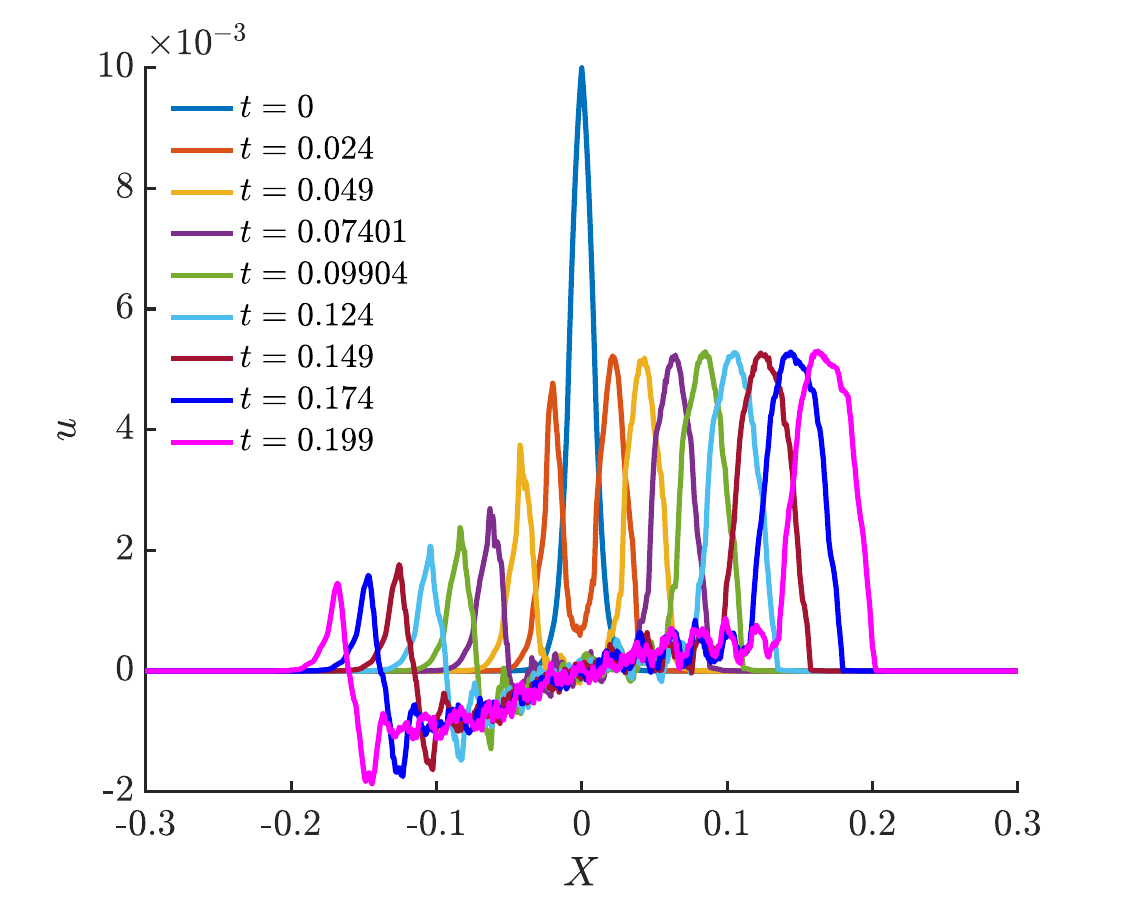}}
    \subfloat[][]{
    \includegraphics[scale=0.45]{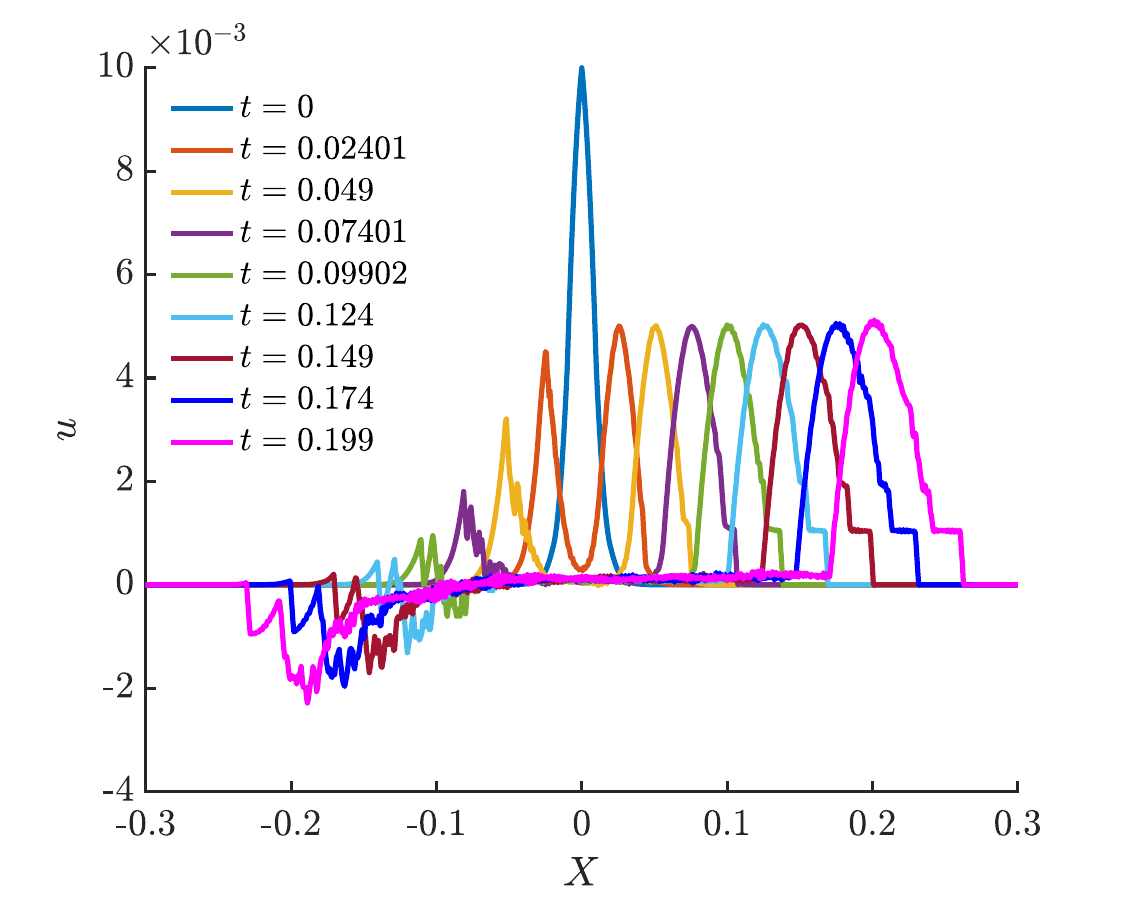}}
    \caption{Effect of initial displacement profile on wave propagation using $a=0.01$, $c= 0.01$ for (a) heterogeneous case $(C = 0.5)$, and (b) homogeneous case $(C = 1.0)$, in VME simulations.}
    \label{fig:vme_dns_initial_disp_shape_effects}
\end{figure}

Figure~\ref{fig:vme_dns_initial_disp_shape_effects} shows the results obtained from the VME simulations for both heterogeneous and homogeneous problems. The relative error in the displacement field compared to DNS at $t=0.199$ is $0.0976$ for the heterogeneous case, and $0.2226$ for the homogeneous case. Since the initial wavelength is small, it is observed that the propagating wave interacts with the heterogeneous microstructure, resulting in oscillations as evident in Fig.~\ref{fig:vme_dns_initial_disp_shape_effects}a. In the homogeneous case, the oscillations are relatively small and are primarily caused by the variation in the local wave speed, unlike the heterogeneous problem.

\subsection{Computational cost comparison for integration schemes} \label{sec:cost_comparison_integration_schemes}
We compare the computation time for different time integration schemes in VME for different contrast values in the elastic modulus for a heterogeneous microstructure. Table \ref{tab:cost_comparison} reports the run time in seconds and the relative error with respect to DNS for various contrast values in elastic modulus and corresponding initial displacement condition (by setting $a$ and $c$ in Eq.~\eqref{eq:initial_displacement_1d}) up to non-dimensional time $t = 0.2$ using EE-SSM and EI-SSM integration schemes. The size of the microstructure in all cases is $l = 0.01$, and each unit cell is mapped to a single coarse-scale element i.e., $n_{\mathrm{ecp}} = 1$. The $\mathrm{CFL}$ values are chosen to obtain stable evolution for the multiscale problem, and the relative errors are reported for different $\mathrm{CFL}$, specifically for EI-SSM-based time integration to demonstrate the relative gain in the accuracy at the expense of compute time. 

The reported computational times are obtained from a single-processor implementation and are higher than those of DNS, since the multiscale coupled problems are solved iteratively at each time increment. The formulation, however, naturally lends itself to parallelization, as the uncoupled fine-scale problems for individual unit cells can be computed independently, thereby enabling a reduction in overall runtime. Without model order reduction, the VME approach involves essentially the same number of degrees of freedom as DNS. Nevertheless, its structure provides a foundation for more efficient methods in which the fine-scale problem can be replaced by a data-driven surrogate model.

\begin{table}[htb!]
\small
\begin{center}
\begin{tabular}{|c|c|c|c| c| c| c|} 
 \hline
 \multirow{2}{*}{Contrast} & Initial & Integration  & \multirow{2}{*}{$n_{\mathrm{ef}}$} & \multirow{2}{*}{$\mathrm{CFL}$} & Compute time & Relative  \\
 & displacement & method &  &  & (secs) & error  \\
 \hline
 \multirow{3}{*}{$C = 1.0$} & \multirow{3}{*}{$a= 0.04$, $c=0.05$} &  EE-SSM & \multirow{3}{*}{$8$} & 1.0 & 264.03 & 0.00187 \\ 
  &  & EI-SSM & & 1.0 & 148.35 & 0.0242  \\ 
   &  &  EI-SSM & & 0.5 & 487.07 & 0.0168 \\ \hline
 \multirow{4}{*}{$C = 0.5$} & \multirow{4}{*}{$a= 0.04$, $c=0.05$}  & EE-SSM & \multirow{3}{*}{$8$}  & 1.0 & 289.49  & 0.0104 \\ 
  &  & EI-SSM & & 1.0 & 174.02 & 0.0806 \\
  &  & EI-SSM & & 0.5 & 550.82 & 0.0588 \\
  &  &  EI-SSM & & 0.25 & 1726.32 &  0.0557 \\ \hline
  \multirow{3}{*}{$C = 0.2$} & \multirow{3}{*}{$a= 0.04$, $c=0.05$} & EE-SSM & $8$  & 0.2 & 1549.77 & 0.07801  \\ 
  &  & EI-SSM & $16$ & 0.5 & 544.48 & 0.3467 \\
  &  & EI-SSM & $16$ & 0.1 &  5171.20 & 0.2014 \\ \hline
  \multirow{2}{*}{$C = 0.01$} & \multirow{2}{*}{$a= 0.005$, $c=0.05$} & EE-SSM & $8$  & 0.1 &  1059.19 & 0.0415 \\ 
  &  &  EI-SSM & $16$ & 0.05 & 6515.87 & 0.2166 \\
 \hline
\end{tabular}
\end{center}
\caption{Comparison of computational cost and relative error with respect to DNS of different time integration methods for varying contrasts in elastic modulus and different $\mathrm{CFL}$ values.}
 \label{tab:cost_comparison}
\end{table}

As shown in Table~\ref{tab:cost_comparison}, for the homogeneous case, the EI-SSM scheme is less expensive than EE-SSM while maintaining a relative error within $2.5 \%$ for EI-SSM. The lower cost of EI-SSM arises because the stable time increment is controlled by the coarse-scale problem, allowing for larger time steps that offset the additional expense of solving nonlinear systems at the fine scale. For the heterogeneous case with moderate stiffness contrast $(C=0.5)$, EI-SSM still permits large stable time increments; however, achieving a relative error within $6\%$ compared to DNS requires smaller time steps. For higher contrasts, $C=0.2$ and $C=0.01$, EI-SSM yields relatively large errors, and EE-SSM outperforms it in both accuracy and computational cost. In fact, even with fewer fine-scale elements, EE-SSM achieves higher accuracy. 

Overall, while the implicit scheme benefits from larger stability limits, it suffers from reduced accuracy. This is because the difference between the stable time increment for explicit evolution of the fine-scale problem and the time increment required for accurate evolution is small, given the second-order accuracy of the current scheme \cite{noh2013explicit, bathe2012insight, bathe2005composite}. This limitation could be alleviated by adopting higher-order integration schemes, which would allow larger implicit time steps while retaining the required accuracy for the multiscale problem. Another way to improve the accuracy of the EI-SSM scheme is to impose a tighter convergence tolerance between the fine- and coarse-scale problems within the operator-split procedure. However, this would also increase the overall computational cost.

\vspace{-0.3cm}
\section{Conclusion} \label{sec:conclusion}
The proposed multiscale computational framework enables the simulation of wave propagation under scale-inseparable conditions, including short-wavelength regimes, while accounting for both material and geometric nonlinearities. By employing an additive decomposition of the solution fields into coarse- and fine-scale components, a coupled two-scale system of equations is derived. The framework allows the discretization of each unit cell with a patch of coarse-scale elements, which is essential to accurately capture wave propagation, especially in short-wavelength regimes. The coarse-scale semi-discrete equations are integrated explicitly, while the fine-scale equations are integrated either explicitly or implicitly, using both dissipative and non-dissipative time integration schemes. 

The numerical examples demonstrate that the VME method accurately captures wave dispersion, attenuation, and wave steepening arising from microstructural heterogeneity, microstructural size, and nonlinearities in the constitutive model, with results in close agreement with DNS. It is shown how the relative error reduces using multiple coarse-scale elements that discretize a unit cell, when the microstructure size is large, leading to wave dispersion due to heterogeneity. It is also observed that the EE-SSM scheme is computationally less expensive than EI-SSM for heterogeneous one-dimensional problems. Although EI-SSM permits larger stability limits, its accuracy is limited by the second-order nature of the integration schemes. In higher-dimensional heterogeneous problems, however, the explicit-implicit approach may provide computational advantages, depending on the relative constraints imposed by stability and accuracy requirements for the fine-scale problem.

Future work will focus on implementing the current framework in higher dimensions for modeling the dynamic response of architected materials. An additional key direction is the development of reduced-order VME models using data-driven surrogate approaches, where the fine-scale problem is replaced with a trained surrogate model, thereby achieving computational efficiency beyond DNS.

\vspace{-0.3cm}
\section*{Acknowledgments}

The authors wish to acknowledge the financial support of the Army Research Office through the Solid Mechanics Program (Grant No. W911NF2320134; Program Officer: Dr. Denise Ford).

\appendix
\renewcommand{\theequation}{\Alph{section}.\arabic{equation}}

\section{Neo-Hookean material} \label{sec:neo-hookean-model}
\setcounter{equation}{0}
The one-dimensional energy density function for the Neo-Hookean material model is obtained from the multi-dimensional form shown below:
\begin{equation}
    \psi (\bfF) = \frac{\mu}{2} \Big[ \ttrace (\bfF^T \cdot \bfF) - n_\mathrm{dim} \Big] - \mu \ln (J) + \frac{\lambda}{2} \ln^2 (J),
\end{equation}
where $J$ denotes the Jacobian of deformation gradient tensor $J = \text{det} (\bfF)$, $\mu$ and $\lambda$ are Lame's parameters, and $n_\mathrm{dim}$ denotes the number of spatial dimensions. The first Piola-Kirchhoff stress for the multi-dimensional case is given below:
\begin{equation}
    \bfP = \frac{\partial \psi (\bfF)}{ \partial \bfF} = \Big[ \lambda \ln (J) - \mu \Big] \bfF^{-T} + \mu \bfF.
\end{equation}
In the one-dimensional case, the deformation gradient tensor can be written as:
\begin{equation}
    \bfF = \left( 1+ \frac{\partial u}{\partial X} \right) \bfe_{1} \otimes \bfe_{1} + \bfe_{2} \otimes \bfe_{2} + \bfe_{3} \otimes \bfe_{3}.
\end{equation}
Furthermore, assuming Poisson's ratio to be $\nu = 0$, yields the Lam\'e parameters in the model to be $\lambda = 0$ and $\mu = E/2$. Denoting $F:= 1 +  \frac{\partial u}{\partial X}$, the energy density function and the only non-zero component of the first Piola-Kirchhoff stress tensor, denoted as $P$, in the 1-D case are obtained as given below:
\begin{equation}
\begin{aligned}
    & \psi (F) = \frac{E}{4} \left(\left(1 + \frac{\partial u}{\partial X} \right)^2 - 1 - 2 \ln{ \left( 1+ \frac{\partial u}{\partial X} \right) } \right),  \\
    & P = \frac{\partial \psi}{ \partial F} \equiv \frac{E}{2} \left(  1+ \frac{\partial u}{\partial X} - \left(1 + \frac{\partial u}{\partial X} \right)^{-1} \right).
\end{aligned}
\end{equation}
\section{Estimate of critical time step in 1-D: DNS} \label{sec:dns_eigenvalue}
\setcounter{equation}{0}

The element-level eigenvalue problem for the DNS is obtained by taking the variation of the weak form of the governing equations given in Eq.~\eqref{eq:weak_form_onescale}. In the 1-D case, the first variation along the direction $(\tilde{d}u, \tilde{d}\ddot{u})$ of the weak form with the Neo-Hookean material (ignoring the forcing terms), is obtained as given below:
\begin{equation}
     \int_{\Omega} \delta u \: \rho_0 \: \frac{\partial^2 (\tilde{d} u)}{\partial t^2} \: dX + \int_{\Omega} \frac{\partial \delta u}{\partial X} \frac{E}{2} \left( 1 + \frac{1}{\left( \partial x / \partial X\right)^2} \right) \frac{\partial \tilde{d} u}{ \partial X} \: dX  = 0,
\end{equation}
where the current displacement field is $u$, and the current deformed configuration is $x = X+u$. The element-level mass matrix and linearized stiffness matrix are given by:
\begin{equation}
\begin{aligned}
    & \bfK_e = \int_{\Omega_e} \frac{E}{2} \left( 1 + \frac{1}{\left( \partial x / \partial X\right)^2} \right)  (\bfB_e)^T  \bfB_e \: dX, \\
    & \bfM_e =  \int_{\Omega_e} \rho_0 (\bfN_e)^T \bfN_e \: dX ,
\end{aligned}
\end{equation}
where $\bfN_e$ and $\bfB_e$ are the shape function matrix and shape function gradient matrix for an element $\Omega_e$. The maximum eigenvalue of the element-level problem is bounded by the maximum eigenvalue of the integrands of stiffness and mass matrices at each quadrature point \cite{belytschko1985stability, belytschko2014nonlinear}. For a quadratic element with a diagonal mass matrix by row-sum lumping, the maximum element-level eigenvalue is estimated to be \cite{belytschko2014nonlinear}:
\begin{equation} \label{eq:wave_speed}
     (\omega_0)_e =  \frac{2 \sqrt{6}}{h_e} \max_{\zeta_{Q,e}} \left( \sqrt{\frac{E}{2 \rho_0} \left( 1 +  \frac{1}{\left( \partial x / \partial X\right)^2} \right)} \right),
\end{equation}
where $\max(\cdot)$ denotes the maximum over all the quadrature points in an element $(\zeta_{Q, e})$, and $h_e$ denotes the length of an element. The critical time step is then obtained by taking the maximum eigenvalue over all elements as given below:
\begin{equation}
    \Delta t_{\mathrm{crit}} = \mathrm{CFL} \max_e \frac{2}{(\omega_0)^e} = \mathrm{CFL} \min_e \left( \frac{h_e}{\sqrt{6}} \min_{\zeta_{Q,e}} \left( \frac{1}{\sqrt{\frac{E}{2 \rho_0} \left( 1 +  \frac{1}{\left( \partial x / \partial X\right)^2} \right)}} \right)\right).
\end{equation}
As noted in \cite{belytschko2014nonlinear}, these estimates are good for problems with $C^1$ constitutive laws and smooth response, and for rough problems like impact, reductions in time steps are advised by appropriately choosing smaller values of $\mathrm{CFL}$. 
\section{Estimate of critical time step in 1-D: multiscale problem} \label{sec:vme_eigenvalue}
\setcounter{equation}{0}
The critical element-level eigenvalue problems for the coarse and fine-scale problems in VME are obtained by taking the first variation of the weak form of the corresponding equations. Following the procedure discussed in Appendix \ref{sec:dns_eigenvalue}, one can obtain maximum element-level eigenvalues for fine and coarse-scale elements for a 1-D problem as given below:
\begin{equation}
\begin{aligned}
    & (\omega^{\mathrm{f}_\alpha}_0)_{e} = \frac{2 \sqrt{6}}{h^{\mathrm{f}_\alpha}_e} \max_{\zeta_{Q,e}} \left( \sqrt{\frac{E}{2 \rho_0} \left( 1 +  \frac{1}{\left( \partial x / \partial X\right)^2} \right)} \right), \\
    & (\omega^{\mathrm{c}}_0)_{\alpha, E} = \frac{2 \sqrt{6}}{h^{\mathrm{c}}_{\alpha, E}} \max_{\zeta_{Q,\alpha, E}} \left( \sqrt{\frac{E}{2 \rho_0} \left( 1 +  \frac{1}{\left( \partial x / \partial X\right)^2} \right)} \right),
\end{aligned}    
\end{equation}
where $(\omega^{\mathrm{f}_\alpha}_0)_{e}$ and $(\omega^{\mathrm{c}}_0)_{\alpha, E}$ are the fine and coarse-scale maximum eigenvalues for the corresponding fine-scale element ($e$) and coarse-scale element $(\alpha, E)$. $h^{\mathrm{f}_\alpha}_e$ and $h^\mathrm{c}_{\alpha, E}$ are the element lengths for fine and coarse-scale elements, respectively. It must be noted that, for the coarse-scale estimate above, the maximum is defined over all the quadrature points in all fine-scale elements $(e)$ within a coarse-scale element $(\alpha, E)$.
Finally, the critical time steps for fine and coarse-scale problems are given by:
\begin{equation}
\begin{aligned}
    & \Delta t^{\mathrm{f}_\alpha}_{\mathrm{crit}} = \mathrm{CFL} \max_e \frac{2}{(\omega^{\mathrm{f}_\alpha}_0)_{e}} = \mathrm{CFL} \min_e \left( \frac{h^{\mathrm{f}_\alpha}_e}{\sqrt{6}} \min_{\zeta_{Q,e}} \left( \frac{1}{\sqrt{\frac{E}{2 \rho_0} \left( 1 +  \frac{1}{\left( \partial x / \partial X\right)^2} \right)}} \right)\right), \\
    & \Delta t^\mathrm{c}_{\mathrm{crit}} = \mathrm{CFL} \max_{\alpha, E} \frac{2}{(\omega^{\mathrm{c}}_0)_{\alpha, E}} = \mathrm{CFL} \min_{\alpha, E} \left( \frac{h^\mathrm{c}_{\alpha, E}}{\sqrt{6}} \min_{\zeta_{Q,\alpha, E}} \left( \frac{1}{\sqrt{\frac{E}{2 \rho_0} \left( 1 +  \frac{1}{\left( \partial x / \partial X\right)^2} \right)}} \right)\right).
\end{aligned}
\end{equation}

{\footnotesize \bibliography{References}}

\end{document}